\documentclass[%
reprint,
superscriptaddress,
nofootinbib,
amsmath,amssymb,
aps,
prb,
floatfix,
]{revtex4-2}
\usepackage{amsmath,amssymb}
\usepackage{graphicx}% Include figure files
\usepackage{dcolumn}% Align table columns on decimal point
\usepackage{bm}% bold math
\usepackage{lineno}
\usepackage{xcolor}
\usepackage{hyperref}% add hypertext capabilities
\begin{document}	

\title{Multiple truly topological unidirectional surface magnetoplasmons at terahertz frequencies}% Force line breaks with \\%

\author{Shengquan Fan}
\thanks{These authors contributed equally to this work.}
\affiliation{School of Physical and Material, Nanchang University, Nanchang 330031, China}
\affiliation{Institute of Space Science and Technology, Nanchang University, Nanchang 330031, China}

\author{Tianjing Guo}
\thanks{These authors contributed equally to this work.}
\affiliation{Institute of Space Science and Technology, Nanchang University, Nanchang 330031, China}

\author{Binbin Zhou}
\affiliation{Department of Electrical and Photonics Engineering, Technical University of Denmark, Kongens Lyngby DK-2800, Denmark}

\author{Jie Xu}
\affiliation{School of Medical Information and Engineering, Southwest Medical University, Luzhou 646000, China}

\author{Xiaohua Deng}
\affiliation{Institute of Space Science and Technology, Nanchang University, Nanchang 330031, China}

\author{Jiangtao Lei}
\affiliation{Institute of Space Science and Technology, Nanchang University, Nanchang 330031, China}

\author{Yun Shen}
\affiliation{School of Physical and Material, Nanchang University, Nanchang 330031, China}

\author{Meicheng Fu}
\email{fumeicheng10@nudt.edu.cn}
\affiliation{College of Science, National University of Defense Technology, Changsha 410073, China}

\author{Kosmas L. Tsakmakidis}
\email{ktsakmakidis@phys.uoa.gr}
\affiliation{Section of Condensed Matter Physics, Department of Physics, National and Kapodistrian University of Athens Panepistimioupolis, Athens GR-157 84, Greece}

\author{Lujun Hong }
\email{ljhong@ncu.edu.cn}
\affiliation{Institute of Space Science and Technology, Nanchang University, Nanchang 330031, China}

\date{\today}% It is always \today, today,
%  but any date may be explicitly specified
\begin{abstract}
Unidirectional propagation based on surface magnetoplasmons (SMPs) has recently been realized at the interface of magnetized semiconductors. However, usually SMPs lose their unidirectionality due to non-local effects, especially in the lower trivial bandgap of such structures. More recently, a truly unidirectional SMP (USMP) has been demonstrated in the upper topological non-trivial bandgap, but it supports only a single USMP, limiting its functionality. In this work, we present a fundamental physical model for multiple, robust, truly topological USMP modes at terahertz (THz) frequencies, realized in a semiconductor-dielectric-semiconductor (SDS) slab waveguide under opposing external magnetic fields. We analytically derive the dispersion properties of the SMPs and perform numerical analysis in both local and non-local models. Our results show that the SDS waveguide supports two truly (even and odd) USMP modes in the upper topological non-trivial bandgap. Exploiting these two modes, we demonstrate unidirectional SMP multimode interference (USMMI), being highly robust and immune to backscattering, overcoming the back-reflection issue in conventional bidirectional waveguides. To demonstrate the usefulness of this approach, we numerically realize a frequency- and magnetically-tunable arbitrary-ratio splitter based on this robust USMMI, enabling multimode conversion. We, further, identify a unique index-near-zero (INZ) odd USMP mode in the SDS waveguide, distinct from conventional semiconductor-dielectric-metal waveguides. Leveraging this INZ mode, we achieve phase modulation with a phase shift from -$\pi$ to $\pi$. Our work expands the manipulation of topological waves and enriches the field of truly non-reciprocal topological physics for practical device applications.
\end{abstract}

%\keywords{Suggested keywords}%Use showkeys class option if keyword
%display desired

%\tableofcontents

\maketitle

\section{Introduction } 
Topological electromagnetics (EM) \cite{lutop2014NP,huBrokenTime2024NC,pruTopPhases2022PRL,RosiekBackscattering2023NP} has gained significant attention due to its intriguing physics and potential applications \cite{BahariNonTop2017SCi}. One of the most remarkable features is the existence of unidirectional EM edge modes \cite{GaoUnidirectionalBulk2023PR,yuTopLargeareaOneway2023PR} in non-trivial topological bandgaps, which can propagate in a single direction while suppressing backward reflection, even in the presence of defects \cite{ao2009onePRB,AleTopRobust2024NC}. These unidirectional edge modes arise from the breaking of time-reversal symmetry through an external magnetic field \cite{HadadMagnetized2010PRL,ChenPreRobust2022PRL}, and were first proposed as analogues of quantum Hall edge states \cite{haldane2008PRL,Prange} in photonic crystals (PhCs) \cite{Joannopoulos}. Such PhCs-based unidirectional modes have been demonstrated both theoretically \cite{wang2008PRL,LuTopOneFiber2018NC,FANG2024110309} and experimentally \cite{wang2009nature,PooExpGuiding2011PRL,WangTopLarge2021PRL,ZhouObserve2020PRL,ChenMultiple2024PRL,liuAxion2025SCi} at microwave frequencies.

As an alternative, surface magnetoplasmons (SMPs) have been proposed for unidirectional propagation due to their simple and robust structure \cite{yu2008onePRL,hu2012broadly,GaoPhotonic2016NC,tsakmakidisTopMulti2021APL,FerTopEle2019PRA, OzawaTopPhotonics2019RMP}. Recently, we have realized unidirectional SMPs (USMPs) using gyromagnetic and gyroelectric materials \cite{ShenlGuidedwave2024NC,ZhouINz2022Oe,TsaowWave2017Sci,XuTopHig2023CP,tsakmakidis2017Sci}. In the terahertz (THz) regime, two types of USMP have been identified \cite{jintopmagne2016NC,hassani2019truly}. The first type, non-topological USMP, exists in the lower trivial bandgap of a magnetized semiconductor and transparent dielectric ($\epsilon_r>0$) waveguide \cite{tsakmakidis2017Sci,brion1972theory,tsakmakidisNonTimeband2020OP,has2020PRA}, maintaining unidirectional characteristics due to non-reciprocal flat asymptotes \cite{tsakmakidis2017Sci}.
However, these SMPs lose their strict unidirectionality when non-local effects are included \cite{raza2015nonlocal}, as the asymptotes vanish \cite{buddhiraju2020absence}. The second type, truly topological USMPs, exist in the upper non-trivial bandgap \cite{hassani2019truly, TopWave2018PRL,CouTop2018AWPL,BassToplExt2022PRR,MonticoneTruly2020NP}, 
characterized by a nonzero gap Chern number ($C_{gap}$) \cite{Continuous2015PRB},
at the interface between magnetized semiconductors and opaque dielectrics ($\epsilon_r<0$), and are immune to non-local effects. These genuinely topological USMPs were first theoretically proposed and shown to exhibit robust unidirectionality against non-locality \cite{hassani2019truly}, and later experimentally demonstrated in a magnetized InSb-metal waveguide \cite{LiangGyrotropic2021OP}. Further, a low-loss broadband USMP in the upper bandgap was proposed in an InSb-Si-air-metal waveguide \cite{yan2023broadband}, though only one truly USMP mode is supported in these waveguides. 

Moreover, multimode interference (MMI) has been realized by coupling two or more conventional edge modes\cite{liu2022multimode,MaTHzDevices2024LT}, but these modes are bidirectional and suffer from back reflection. To overcome this limitation, PhCs-based MMI using multiple unidirectional modes has been proposed, effectively suppressing backscattering \cite{Liu2023LT,SkiMultimode2014PRL}. More recently, unidirectional MMI has been experimentally demonstrated in a PhCs waveguide under two external magnetic fields at microwave frequencies \cite{tang2024magnetically}. Using USMP, microwave MMI and mode conversion is achieved by applying opposing magnetic ﬁelds
 \cite{liu2025OL}. However, unidirectional MMI and mode conversion based on USMPs has not been reported at THz frequencies.

In this work, we present a fundamental physical model for multiple truly USMP modes at THz frequencies, differing from previously proposed waveguides that support only one true USMP mode \cite{hassani2019truly, LiangGyrotropic2021OP, yan2023broadband}. The model involves a semiconductor-dielectric-semiconductor (SDS) waveguide under opposing magnetic fields. We analytically derive and numerically analyze the dispersion in both local and non-local models, demonstrating that this waveguide supports two truly even and odd USMP modes in the upper topological bulk mode bandgap while these modes lose their unidirectionality in the lower bulk mode bandgap due to non-local effects. Using these two modes, we realize unidirectional SMP multimode interference (USMMI), which exhibits strong robustness against defects and eliminates backscattering, effectively overcoming the backscattering problem in conventional bidirectional waveguides; similarly to the PhCs-based MMI at microwave frequencies \cite{tang2024magnetically}.  Additionally, we demonstrate a frequency- and magnetically-tunable arbitrary-ratio power splitter based on this robust USMMI, along with mode conversion capabilities. Notably, we report a unique index-near-zero (INZ) odd USMP mode in our waveguide, contrasting with conventional semiconductor-dielectric-metal waveguides that do not support the INZ mode \cite{tsakmakidis2017Sci}. This discovery enables 
efficient phase modulation with a phase shift of $2\pi$.

\section{ Theoretical Physical model }
The basic physical model of multiple truly THz USMP modes in the SDS waveguide is illustrated in Fig.~\ref{fig:epsart}. In this part, we theoretically derive the dispersion equations for the SMP supported by the SDS structure in both local and non-local models.  
\subsection{\label{sec:level1} Dispersion of SMP in the local model}
First, we derive the dispersion equation of SMP in the local model. In our SDS system, the dielectric constant and thickness of the dielectric layer are $\varepsilon_r$ and $2d$, respectively. To break the time-reversal symmetry of the system, opposing external magnetic fields ($B_1$ and $B_2$) are imposed on the semiconductors in the $z$ direction. Consequently, the semiconductors exhibit gyroelectric anisotropy, with two corresponding relative permittivity tensors \cite{tsakmakidis2017Sci,brion1972theory} 
\begin{align}
	&{\varepsilon_{s_1}}=\begin{bmatrix}
		{\varepsilon_{1_1}} & \mathrm{i}{\varepsilon_{2_1}}& 0 \\
		-\mathrm{i}{\varepsilon_{2_1}} & {\varepsilon_{1_1}} & 0\\
		0 & 0 &  \varepsilon_{3}  
	\end{bmatrix},
	{\varepsilon_{s_2}}=\begin{bmatrix}
		{\varepsilon_{1_2}} & -\mathrm{i}{\varepsilon_{2_2}} & 0 \\
		\mathrm{i}{\varepsilon_{2_2}}& {\varepsilon_{1_2}} & 0\\
		0 & 0 &  \varepsilon_{3}
	\end{bmatrix} 
\end{align}
with
${\varepsilon_{1_j}}=\varepsilon_\infty(1-\displaystyle\frac{\overline{\omega}\omega_p^2}{\omega[\overline{\omega}^2-\omega_{c_j}^2]})$, ${\varepsilon_{2_j}}=\varepsilon_\infty\displaystyle\frac{\omega_{c_j}\omega_p^2}{{\omega}(\overline{\omega}^2-\omega_{c_j}^2)}$ ($j=1,2$), and
$\varepsilon_3=\varepsilon_\infty(1-\displaystyle\frac{\omega_p^2}{\overline{\omega}^2})$, where $\overline{\omega}=\omega+i\nu$, $\nu$ is the electron scattering frequency, $\omega_{c_j}=e B_j/ m^*$ (where $e$ and $m^*$ are respectively the charge and effective mass of the electron) is the electron cyclotron frequency, $\omega_p$  is the plasma frequency, and  $\varepsilon_\infty$ is the high-frequency relative permittivity. Note that without an external magnetic field, it is a conventional isotropic material ($\varepsilon_{2_j}=0$ and $\varepsilon_{1_j}=\varepsilon_{3}$).  In the magnetized semiconductor, the bulk modes have a dispersion relation of 
\begin{eqnarray}
	k =\sqrt{\varepsilon_{v_j}}k_0 
\end{eqnarray}
for transverse-magnetic (TM) polarization, where $k$ is the propagation constant, $k_0=\omega/c$ is the vacuum wavenumber, and $\varepsilon_{v_j}=\varepsilon_{1_j}-\varepsilon_{2_j}/\varepsilon_{1_j}^2$ is Voigt permittivity. It has two bandgaps with $\varepsilon_{v_j}<0$. The upper boundaries of the lower and upper bandgaps are $\omega_{a_j} =\sqrt{\omega_{c_j}^2/4+\omega_{p}^2}-\omega_{c_j}/2$ and  $\omega_{b_j}  =\sqrt{\omega_{c_j}^2/4+\omega_{p}^2}+\omega_{c_j}/2$ by $k=0$ into Eq. (2).  The lower boundary of the upper bandgaps are $\omega_{r_j} =\sqrt{\omega_{c_j}^2+\omega_{p}^2}$ by $k\to \pm\infty$ into Eq. (2). The waveguide supports TM polarized SMP. Solving Maxwell’s equations with four continuous boundary conditions, the dispersion relation of SMP in the local model can be derived as (the details see Appendix A)
\begin{eqnarray}
	e^{4\alpha_d d}=\prod_{j=1}^{2} M_j \label{subeq:3}
\end{eqnarray}
with
\begin{align*}
	M_j	=\frac{\varepsilon_{1_j}\alpha_d\varepsilon_{v_j}- \varepsilon_{r}(\varepsilon_{1_j}\alpha_j-k\varepsilon_{2_j})}{\varepsilon_{1_j}\alpha_d\varepsilon_{v_j}+\varepsilon_{r}(\varepsilon_{1_j}\alpha_j-k\varepsilon_{2_j})}
\end{align*}
where $\alpha_d=\sqrt{k^2-\varepsilon_{r} k_{0}^2}$, $\alpha_{j}=\sqrt{k^2-\varepsilon_{v_j} k_{0}^2}$. By substituting $k\to \pm\infty$ into Eq.~(\ref{subeq:3}), the forward ($+$) and backward ($-$) asymptotic frequencies are obtained as $\omega_{sp1}^\pm=\displaystyle\frac{1}{2}(\sqrt{\omega_{c_1}^2+4 \omega_{p}^2 \frac{\varepsilon_\infty}{\varepsilon_\infty +\varepsilon_r}}\mp\omega_{c_1})$ and $\omega_{sp2}^\pm=\displaystyle\frac{1}{2}(\sqrt{\omega_{c_2}^2+4 \omega_{p}^2 \frac{\varepsilon_\infty}{\varepsilon_\infty +\varepsilon_r}}\mp\omega_{c_2})$

Consider a special symmetric case of $B_1 = B_2$ (corresponding to $\omega_{c_1} = \omega_{c_2}$), we have $\varepsilon_{1_1}=\varepsilon_{1_2}=\varepsilon_{1}$, $\varepsilon_{2_1}=\varepsilon_{2_2}=\varepsilon_{2}$, $\varepsilon_{v_1}=\varepsilon_{v_2}=\varepsilon_{v}$ and $\alpha_1=\alpha_2=\alpha$, thus $M_1=M_2$, then Eq.~(\ref{subeq:3}) becomes 
\begin{subequations}
	\begin{eqnarray}
	\label{subeq:4a}\alpha-k \frac{\varepsilon_2}{\varepsilon_1} +\frac{\varepsilon_v}{\varepsilon_r}\alpha_d tanh(\alpha_d d)=0\\ 
		\alpha-k \frac{\varepsilon_2}{\varepsilon_1}+\frac{\varepsilon_v}{\varepsilon_r}\alpha_d coth(\alpha_d d) =0 
	\end{eqnarray} \label{subeq:4}
\end{subequations}
for the even and odd modes, respectively.

\subsection{\label{sec:level2} Dispersion of SMP in the non-local model}
We further derive the dispersion equation of SMP in the non-local model. The non-local response in plasmonic materials originates from the convective and diffusive motion of free electrons during an optical cycle. Several non-local models have been proposed, including the hydrodynamic model \cite{cir2012SCi,HyModels2018ATS,FonFirstTop2024PRR}, random phase approximation \cite{feibel1974PRB}, and a quantum-corrected model \cite{Quan2012NC}. Among these, the hydrodynamic model based on free electron gas \cite{HalHyd1995PRB} has been widely applied in plasmonic materials, such as deep subwavelength metal \cite{Metal2021MNA, Stamatopoulou:22} and doped semiconductors (particularly n-type InSb) \cite{Size2017EPL}. Here, we investigate the nonlocal effects in the SDS system using the hydrodynamic model of the free electron gas, as recently reported in refs. \cite{hassani2019truly, buddhiraju2020absence}. The response of the gyroelectric semiconductors to EM field gives rise to an induced free electron current ${\bf J}$ \cite{hassani2019truly, buddhiraju2020absence,raza2015nonlocal}, which satisfies the hydrodynamic equation $\beta^2\nabla(\nabla\cdot \textbf{J})+\omega(\omega+i\nu)\textbf{J}+i\omega\textbf{J} \times\omega_{c} \hat z=i\omega\omega_p^2\varepsilon_{0}\varepsilon_\infty \textbf{E}$, where $\beta$ is the non-local parameter, and $\nu$ is a phenomenological damping rate. Note that $\beta$ is proportional to the Fermi velocity $\nu_{F}$ ($\beta^2=3\nu_{F}^2/5$ for $\nu=0$), which is inversely proportional to the effective mass of electrons $m^*$ \cite{raza2015nonlocal}.
 Maxwell's equations can be expressed as $\nabla\times \textbf{H}=-i\omega\varepsilon_0 \varepsilon_\infty\textbf{E} +\textbf{J}$ and $\nabla\times \textbf{E}=i\omega\mu_0 \textbf{H}$ due to non-local effects. The dispersion relation of bulk modes can be found as \cite{hassani2019truly, yan2023broadband}
\begin{eqnarray}
	Qk^4+D_jk^2+F_j=0 \label{subeq:5}
\end{eqnarray}
where  $Q\!\!=\!\!\beta^2\omega\overline{\omega}$,
$D_{j}\!\!=\!\!(\beta^2\varepsilon_\infty k_0^2+\omega\overline{\omega})(\omega_p^2-\omega\overline{\omega})+\omega^2\omega_{cj}^2$, and $F_{j}\!\!=\!\!\varepsilon_\infty k_{0}^2(\omega\overline{\omega}-\omega_p^2)^2-\varepsilon_\infty k_0^2\omega^2\omega_{cj}^2$. By solving Eq.~(\ref{subeq:5}), we obtain two dispersion relations for the upper and lower bulk modes: $k_{aj}^{2}=(-D_{j}+\sqrt{D_{j}^2-4QF_{j}})/2Q$ and $k_{bj}^{2}=(-D_{j}-\sqrt{D_{j}^2-4QF_{j}})/2Q$, where the lower cutoff frequencies are identical to $\omega_{a_j}$ and $\omega_{b_j}$ in the local model. Here, $k_{aj}$ and $k_{bj}$ are the propagation constants. In contrast to the local model, the upper cutoff frequency $\omega_{r_j}$ does not exist in the non-local model.

Combining hydrodynamic and Maxwell’s equations with six continuous boundary conditions, the SMP dispersion in the non-local model can be expressed as (the details see Appendix B)
\begin{eqnarray}
	e^{4\alpha_d d}=\prod_{j=1}^{2} N_j \label{subeq:6}
\end{eqnarray}
with
\begin{eqnarray*}
	\!\!N_j=\frac{\varepsilon_\infty\alpha_d/\varepsilon_r(\gamma_{j}s_j\!\!+\!\! ip_{j}s_j')\!\!+\!\!k(\gamma_{j}\!\!+\!\!ip_{j})\!\!+\!\!(k^2\!\!-\!\!k_{0}^2\varepsilon_\infty)(s_j'\!\!-\!\! s_j)}{\varepsilon_\infty\alpha_d/\varepsilon_r(\gamma_{j}s_j\!\!+\!\!ip_{j}s_j')\!\!-\!\!k(\gamma_{j}\!\!+\!\!ip_{j})\!\!-\!\!(k^2\!\!-\!\!k_0^2\varepsilon_\infty)(s_j'\!\!-\!\! s_j)}\!\!
\end{eqnarray*}
where $\gamma_{j}=\sqrt{k^{2}-k_{bj}^{2}}$, $p_{j}=\sqrt{k_{aj}^2-k^{2}}$, and
\begin{align*}
	&s_j=\frac{\omega\overline{\omega}p_{j}^2+i\omega\omega_{cj} kp_{j}+\varepsilon_\infty k_0^2[\beta^2k^2-\omega\overline{\omega}+\omega_p^2]}{\omega\omega_{cj}(k^2-\varepsilon_\infty k_0^2)-i(\omega\overline{\omega}-\varepsilon_\infty\beta^2k_0^2)kp_{j}}\\  
	&s_j'=\frac{-\omega\overline{\omega}\gamma_{j}^2-\omega\omega_{cj} k\gamma_{j}+\varepsilon_\infty k_0^2[\beta^2k^2-\omega\overline{\omega}+\omega_p^2]}{\omega\omega_{cj}(k^2-\varepsilon_\infty k_0^2)+(\omega\overline{\omega}-\varepsilon_\infty\beta^2k_0^2)k \gamma_{j}}\\
\end{align*} 

Consider a special symmetric case, we have $s_1=s_2=s$, $s_1'=s_2'=s'$, $p_{1}=p_{2}=p$ and $\gamma_{1}=\gamma_{2}=\gamma$; thus $N_1=N_2$; Eq.~(\ref{subeq:6}) becomes 
\begin{subequations}
	\begin{align}
		\varepsilon_\infty\alpha_d/\!\varepsilon_r(\gamma s\!\!+\!\!i p s')tanh(\alpha_d d)\!\!+\!\!(\! s \!\!-\!\! s'\!)(k^2\!\!-\!\!k_0^2\varepsilon_\infty)\!\!-\!\!k(\gamma\!\!+\!\!i p)\!\!=\!\!0\\
		%\end{align}
		\varepsilon_\infty\alpha_d/\!\varepsilon_r(\gamma s\!\!+\!\!i p s')coth(\alpha_d d)\!\!+\!\!(\! s \!\!-\!\! s'\!)(k^2\!\!-\!\!k_0^2\varepsilon_\infty)\!\!-\!\!k(\gamma\!\!+\!\!i p)\!\!=\!\!0
	\end{align} \label{subeq:7}
\end{subequations}
for the even and odd modes, respectively. 

\section{Simulation Results}
In this part, we conduct a detailed numerical analysis of the SMP dispersion based on the derived equations in the SDS system and demonstrate many degrees of freedom to manipulate SMP modes, including interference, power, and phase, by full-wave simulation. Throughout the paper, the gyroelectric semiconductor is assumed to be n-doped InSb \cite{hassani2019truly, buddhiraju2020absence} with its typical parameters are $\varepsilon_\infty =15.7$ and $\omega_p=4 \pi \times 10^{12}$ rad/s ($f_{\rm{p}}=2$~THz). Like in refs. \cite{hassani2019truly,buddhiraju2020absence}, we take a non-local parameter $\beta= 1.07\times 10^6$ m/s for non-local hydrodynamic model at room temperature. Note that $\beta= 0$ in the local model. The non-gyroelectric dielectric layer is exemplified by silicon with a relative permittivity of $\varepsilon_{r}=11.68$.

\subsection{ The dispersion in the lower bulk mode bandgap}

We first analyze the dispersion of SMP within the lower bulk mode bandgap. This bandgap ranges from $0$ to $\omega_{a_2}$, where $\omega_{a_2} =\sqrt{\omega_{c_2}^2/4+\omega_{p}^2}-\omega_{c_2}/2$ when $\omega_{c_2}\geq\omega_{c_1}$. The special case of the symmetric ($\omega_{c_2}=\omega_{c_1}$) waveguide is considered. Using Eqs.~(\ref{subeq:4}) and~(\ref{subeq:7}), we numerically calculate the dispersion relations of SMPs in both local and non-local models, respectively. In the calculation, we ignore the effect of material loss (i.e., $\nu$ = 0) and take $d=0.04\lambda_{p}$ as an example. Figure~\ref{fig:epsart}(a) shows the dispersion of SMP in the symmetric waveguides for $\omega_{c_1}=\omega_{c_2}=0.25\omega_p$, corresponding to $B_{1}=B_{2}=0.25$~T. As expected, the SDS structure supports two SMP modes: odd mode (solid red line) and even mode (dashed blue line). These two modes exhibit identical  asymptotic frequencies ($\omega_{sp1}^-=\omega_{sp2}^-$,  $\omega_{sp1}^+=\omega_{sp2}^+$). Moreover, these two modes propagate backward only within the frequency range [$\omega_{sp}^-$,$\omega_{a_2}$], corresponding to [0.6425$\omega_{p}$,0.8828$\omega_{p}$]; As a result, a unidirectional propagation range is formed, as seen in the shaded yellow region in Fig.~\ref{fig:epsart}(a). However, when the non-local effects (non-local parameter $\beta= 1.07\times\ 10^6$ m/s) are taken into account \cite{buddhiraju2020absence,hassani2019truly}, the dispersion relations of the SMP modes differ significantly, as shown in Fig.~\ref{fig:epsart}(b). The dispersion curves increase with $k$, and the forward and backward asymptotic frequencies of both modes vanish at large wave numbers, leading to the disappearance of the unidirectional propagation region within the lower bulk mode bandgap.

To further investigate the phenomenon of unidirectional disappearance induced by non-local effects, we analyze the dispersion in the common case of asymmetric waveguide ($\omega_{c_2}\neq\omega_{c_1}$) using Eqs.~(\ref{subeq:3}) and~(\ref{subeq:6}). As an example, $\omega_{c_1} = 0.15\omega_p$ and $\omega_{c_2} = 0.25\omega_p$, which correspond to $B_{1}=0.15$ T and $B_{2}=0.25$~T, and the other parameters are identical to those of the symmetric waveguide. Figure~\ref{fig:epsart}(c) shows the dispersion curve for the local model. Due to the asymmetric structure, the SMP waveguide also supports even and odd modes. The asymptotic frequencies for the odd modes are $\omega_{sp2}^-$ and $\omega_{sp1}^+$, while those for the even modes are  $\omega_{sp1}^-$ and $\omega_{sp2}^+$. As seen in the yellow shaded region, a unidirectional window based on horizontal asymptotes clearly occurs in [$\omega_{sp1}^+$,$\omega_{a_2}$], corresponding to [0.6859$\omega_{p}$,0.8828$\omega_{p}$]. For comparison, the dispersion curve for the non-local model is numerically calculated using Eq.~(\ref{subeq:6}). Similarly to the symmetric waveguide, Fig.~\ref{fig:epsart}(d) further demonstrates the disappearance of the asymptotic frequency when considering non-local effects, leading to the absence of the USMP window in the asymmetric waveguide. Therefore, the results demonstrate that for both symmetric and asymmetric waveguides, the SMP modes lose their unidirectionality in the lower bulk mode bandgap when non-local effects are considered.

\subsection{The dispersion in the upper bulk mode bandgap}

\begin{figure}[t]
	\centering
	\includegraphics[width=0.98\linewidth]{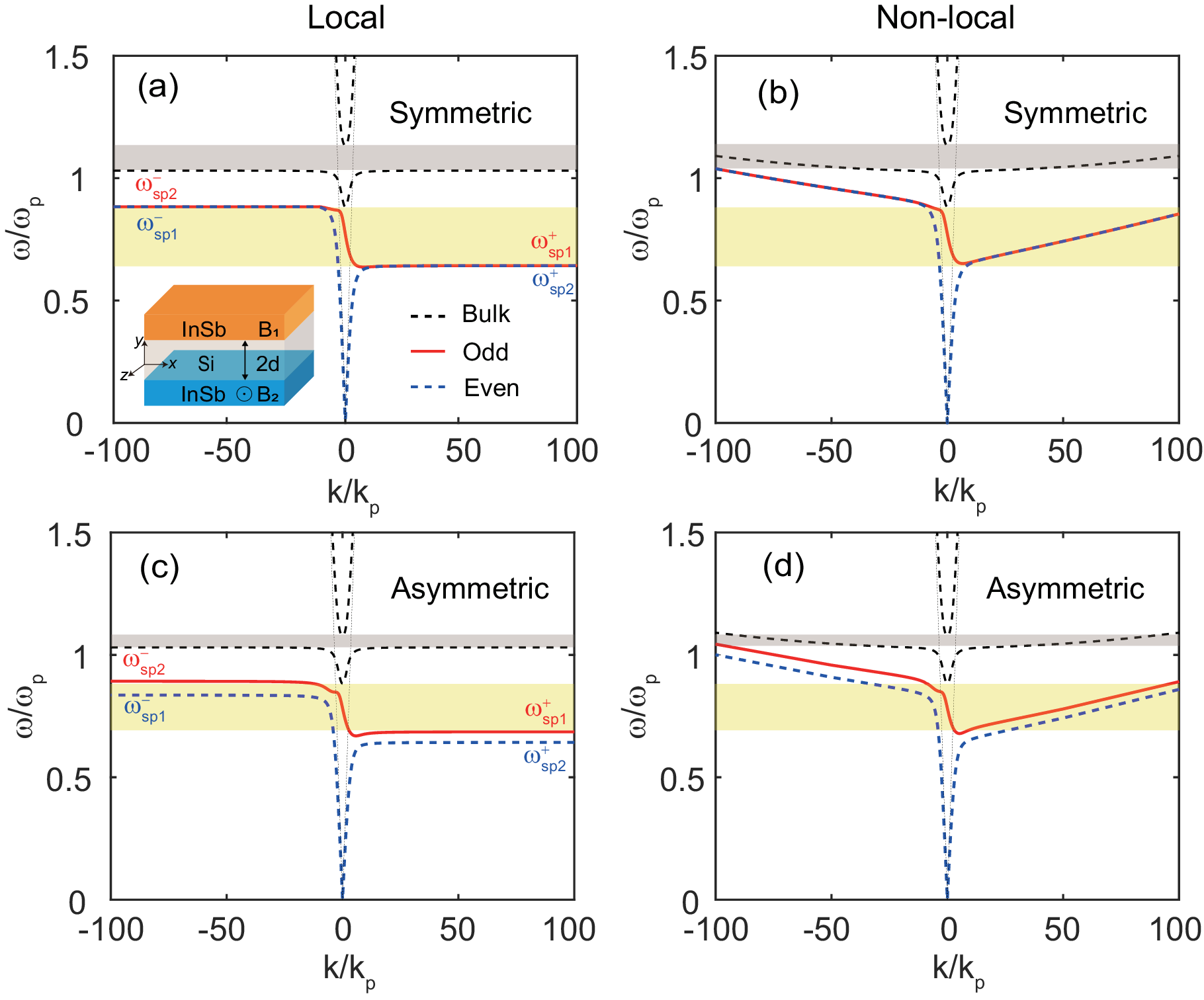}
	\caption{\label{fig:epsart}Dispersion relation of SMP in the lower bulk mode bandgap for the local (a, c) and non-local (b,  d) model. (a,b) show the symmetric waveguide for $\omega_{c_1}=\omega_{c_2}=0.25\omega_p$, while (c, d) show the asymmetric waveguide for $\omega_{c_1} = 0.15\omega_p$ and $\omega_{c_2} = 0.25\omega_p$. The solid red and dashed blue lines show the odd and even SMP modes. The inset shows the schematic of the SDS waveguide. $\omega_{c_1} = 0.15\omega_p$ and $\omega_{c_2} = 0.25\omega_p$, correspond to the external magnetic fields $B_1=0.15$ T and $B_2=0.25$~T, respectively. The non-local parameter is $\beta= 1.07\times\ 10^6$ m/s, and the other parameters are $\varepsilon_\infty =15.7$, $\varepsilon _{r}=11.68$, $d=0.04\lambda_{p}$, and  $\omega_p=4 \pi \times 10^{12}$ rad/s.}
\end{figure}
\begin{figure*}[t]
 \centering
	\includegraphics[width=0.8\linewidth]{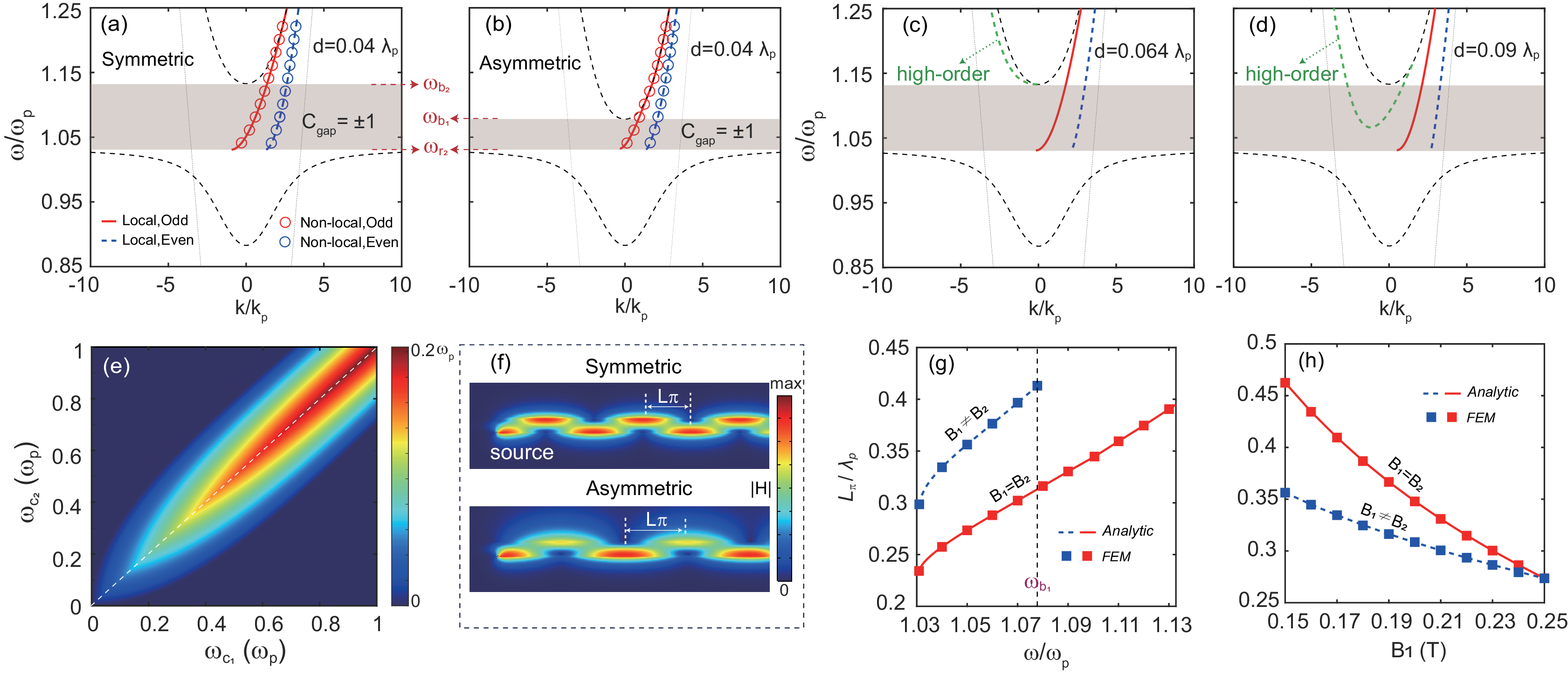}
	\caption{\label{fig:epsart2}(a-d) Dispersion relations of SMPs in the upper bulk mode bandgap. (a,c,d) The symmetric waveguide for $\omega_{c_1}=\omega_{c_2}=0.25\omega_p$, and (b) the asymmetric waveguide for $\omega_{c_1} = 0.15\omega_p$ and $\omega_{c_2} = 0.25\omega_p$, corresponds to $B_{1}=0.15$ T and $B_{2}=0.25$~T. The dispersion curves almost perfectly coincide for the local (lines) and non-local (circles) models. The blue, red, and green show the even, odd, and high-order SMP modes, respectively. The shaded yellow areas represent the USMP band. (a,b) $d=0.04\lambda_{p}$, (c) $d=0.064\lambda_{p}$   and (d) $d=0.09\lambda_{p}$. (e) The USMP bandwidth $\Delta\omega$ as a function of $\omega_{c_1}$ and $\omega_{c_2}$, and the dashed white line corresponds to $\omega_{c_1}=\omega_{c_2}$. (f) $H$-field distributions in the symmetric and asymmetric waveguide at $\omega = 1.05$ $\omega_p$. (g, h) The analytical (line) and FEM (square) results of beat length $L_\pi$ as a function of frequency $\omega$ and external magnetic field $B_1$, respectively. The blue and red represent  $B_1\neq B_2$ and $B_1=B_2$, respectively. The dashed black line corresponds to $\omega_{b_1}= 1.0778\omega_p$. The other parameters are the same as in Fig.~\ref{fig:epsart}.}
\end{figure*}

We further analyze the dispersion of SMP in the upper bulk mode bandgap [$\omega_{r_2}$, $\omega_{b_1}$], where $\omega_{r_2}=\sqrt{\omega_{c_2}^2+\omega_{p}^2}$ and $\omega_{b_1}  =\sqrt{\omega_{c_1}^2/4+\omega_{p}^2}+\omega_{c_1}/2$, corresponding to the gray region in Fig.~\ref{fig:epsart}. We numerically calculate the dispersion of SMP in the local model, and the parameters are consistent with those in Fig.~\ref{fig:epsart}. Figure~\ref{fig:epsart2}(a) shows the dispersion diagram for the symmetric waveguide with $\omega_{c_1}=\omega_{c_2}=0.25\omega_p$. As seen from the lines, this symmetric waveguide supports both odd (red solid line) and even (blue dashed line) USMP modes in upper bulk mode bandgap [$\omega_{r_2}$, $\omega_{b_2}$], which corresponds to [$1.0308\omega_p$,$1.1328\omega_p$], marked by the grey shaded area. For comparison, the dispersion of SMP in the non-local model is also shown as circles in Fig.~\ref{fig:epsart2}(a). Obviously, the dispersion curves for odd and even SMP modes almost completely coincide for both the local and non-local models at small wavenumbers.  This result agrees well with the result that the effect of nonlocality, proportional to $\beta^2k^2$ in Eq. (6), is expected to be small \cite{hassani2019truly}. In contrast to USMP in the lower trivial bandgap, which exists at larger wavenumbers,  the SMP modes maintain their unidirectionality for small wavenumbers in the upper topologically non-trivial bandgap, even when non-local effects are considered. These USMP modes exhibit the nontrivial topological property, which results from the nonzero gap Chern number $C_{gap}=\pm 1$ for the upper and lower magnetized InSb \cite{ChernTop2022PRB}, as shown in Fig. 2(a). The difference in gap Chern number is $\Delta C_{gap}=1-(-1)=2$. Thus, according to the principle of bulk-edge correspondence\cite{BulkedgeCor2016PRB,BulkEdge2020PRL}, the upper bandgap supports two unidirectional topological modes, which is consistent with the odd and even USMP modes.

Figure~\ref{fig:epsart2}(b) illustrates the dispersion curves for the asymmetric waveguide with $\omega_{c_1} = 0.15\omega_p$ and $\omega_{c_2} = 0.25\omega_p$. Similar to the symmetric results in Fig.~\ref{fig:epsart2}(a), the USMP modes still exist, and the lines and circles almost completely overlap. However, the USMP band is compressed to the range [$1.0308\omega_p$,$1.0778\omega_p$], resulting from the asymmetric structure. Note that our waveguide supports high-order SMP mode based on the total internal reflection (TIR) mechanism in the upper bandgap, which needs to be suppressed since they are typically bidirectional. Their dispersion relations in a symmetric waveguide can be obtained by substituting $\alpha_d = ip $ into Eq.~(\ref{subeq:4}), where $p=\sqrt{\varepsilon_{r} k_{0}^2-k^2}$. To suppress the bidirectional high-order mode in the upper bandgap, the thicknesses of the dielectric layer is required to satisfy $\sqrt{\varepsilon_r}k_0 d\leq \pi/2$ at $k=0$ \cite{yan2023broadband}, corresponding to the critical value $d_{c} = \displaystyle\pi c/(2 \sqrt{\varepsilon_{r}} \omega_{b_2})$ with $\omega_{b_2}  =\sqrt{\omega_{c_2}^2/4+\omega_{p}^2}+\omega_{c_2}/2$. For our symmetric waveguide with $\omega_{c_1}=\omega_{c_2}=0.25\omega_p$, it is found that $d_{c}=0.064\lambda_{p}$. To verify this, we calculated the dispersion curves for $d=0.064\lambda_{p}$, and the results are displayed in Fig. 2(c). As expected, the lower cutoff frequency of the higher-order mode is very close to $ \omega_{b_2}$ at $k=0$. In this case, the high-order mode does not influence the USMP band. Figure 2(d) shows the dispersion curves for $d=0.09\lambda_{p}$. Obviously, for $d > d_{c}$, three SMP modes are supported in the upper bandgap, and the USMP band is compressed by the bidirectional higher-order mode. The result demonstrates that the USMP band is signiﬁcantly affected by $d$. Therefore, the thickness parameter of the dielectric layer should be set to $d \leq d_{c}$.

Next, we will analyze the USMP bandwidth of the asymmetric waveguide. When $\omega_{c_2}\geq\omega_{c_1}$,  the USMP bandwidth $\Delta\omega=\omega_{b_1}-\omega_{r_2}$ is characterized by
\begin{eqnarray}
	\Delta\omega=\sqrt{\omega_{c_1}^2/4+\omega_{p}^2}+\omega_{c_1}/2-\sqrt{\omega_{c_2}^2+\omega_{p}^2}
	\label{subeq:8}
\end{eqnarray} 
Note that when $\omega_{c_2} < \omega_{c_1}$, the roles of $\omega_{c_1}$ and $\omega_{c_2}$ in the Eq.~(\ref{subeq:8}) will be interchanged. Figure~\ref{fig:epsart2}(c) illustrates the variation of bandwidth $\Delta\omega$ with respect to  $\omega_{c_1}$ and $\omega_{c_2}$, which correspond to related parameters $B_1$ and $B_2$. Due to $\omega_{c_1}=e B_1/ m^*$ and $\omega_{c_2}=e B_2/ m^*$, the unidirectional bandwidth $\Delta\omega$ is magnetically-controllable by varying $B_1$ and $B_2$. It is found that $\Delta\omega$ exhibits a local maximum when $\omega_{c_1} = \omega_{c_2}$. The white dashed line indicates the distribution of this local maximum, which increases with $\omega_{c_2}$. 

These results confirm that in the upper bandgap, our waveguide supports two truly USMP modes at THz frequencies. The even and odd USMP modes maintain their unidirectionality when considering non-local effects, which is different from the situation in the lower bandgap. Moreover, the unidirectional characteristics of the SMP modes are equivalent in both the local and non-local models. Therefore, we will take the local model in the subsequent numerical calculations and simulations as an example, and our interest focuses on the upper bulk mode bandgap.

\subsection{ Unidirectional MMI based on two USMP modes }
\begin{figure}[t]
	\centering
	\includegraphics[width=0.98\linewidth]{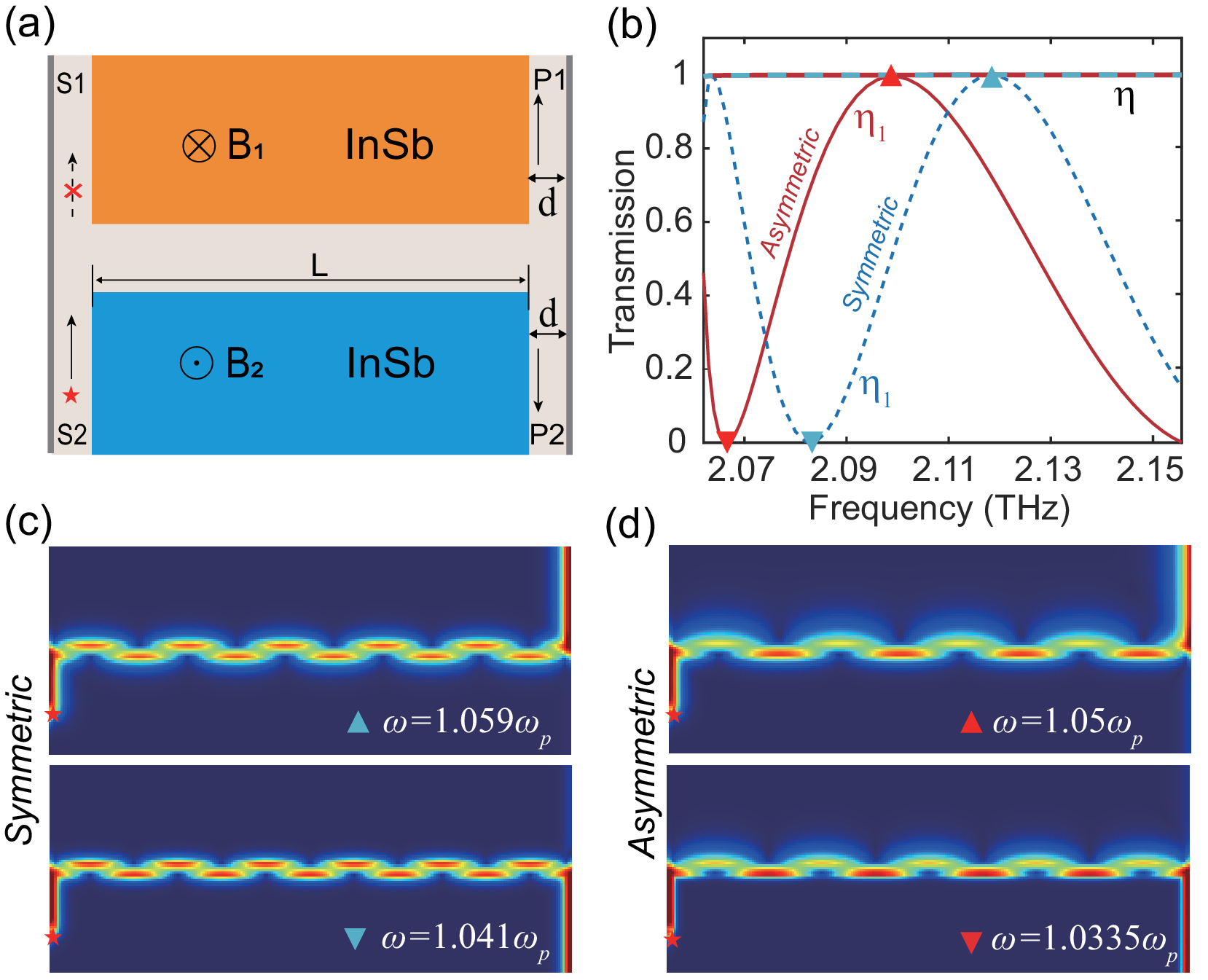}
	\caption{\label{fig:epsart3}Arbitrary ratio power manipulation by frequency. (a) Schematic of the H-shaped splitter based on USMMI. (b) Transmission of SMP along channel P1 ($\eta_1$) and total transmission ($\eta$) in the symmetric and asymmetric splitters. (c,d) Simulated H-field amplitudes in the symmetric splitter at $\omega=1.059\omega_p$ and $\omega=1.041\omega_p$ (c), and the asymmetric splitter at $\omega=1.05\omega_p$ and $\omega=1.0335\omega_p$ (d). The parameters are the same as in Fig.~\ref{fig:epsart2}.}	
\end{figure}
	
\begin{figure}[t]
	\centering
	\includegraphics[width=0.98\linewidth]{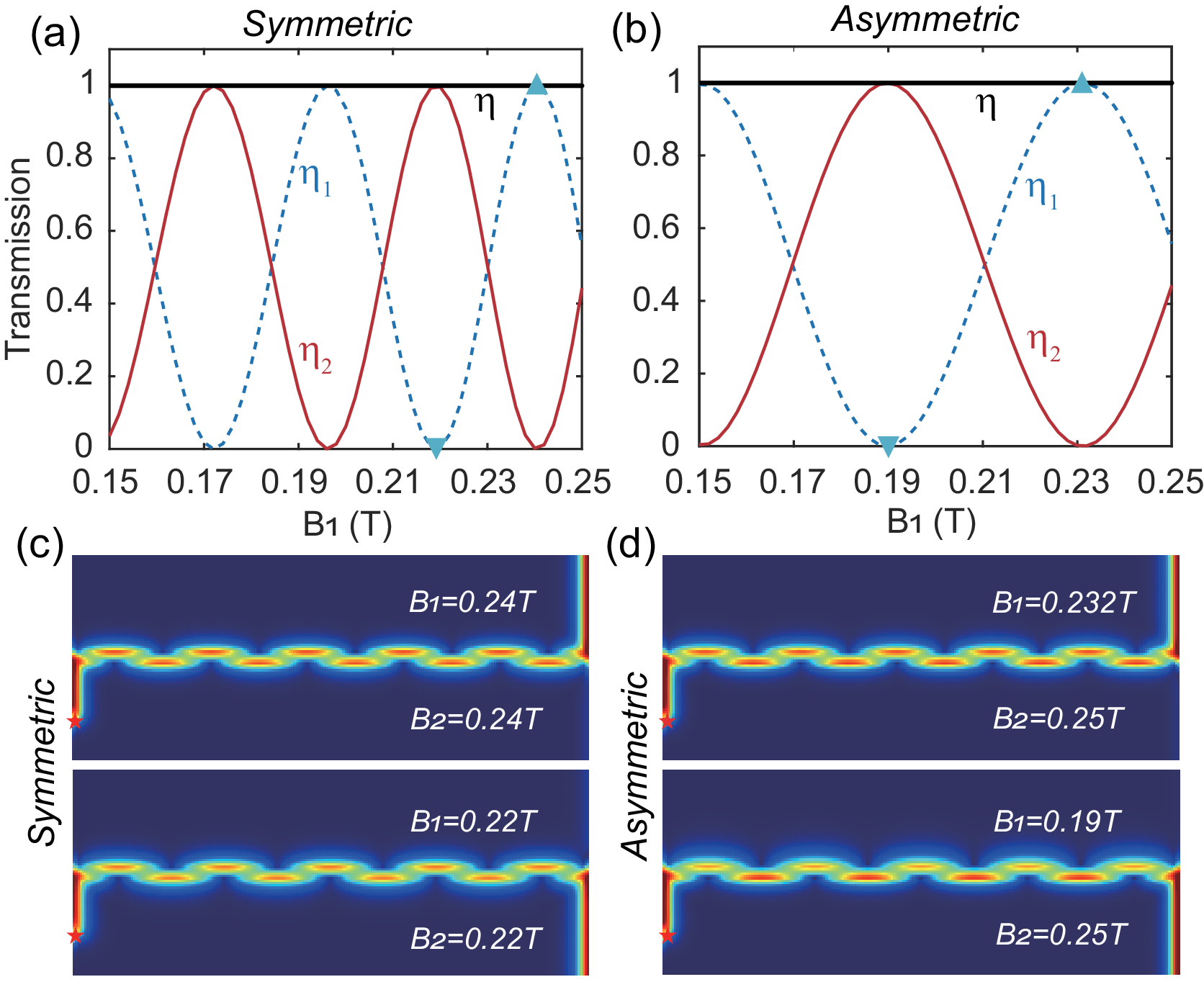}
	\caption{\label{fig:epsart4}Arbitrary ratio power manipulation by the external magnetic field. (a,b) Transmission coefficients $\eta_1$, $\eta_2$, and $\eta$ as functions of magnetic fields $B_1$. (a) $B_1=B_2$, (b) $B_1\neq B_2$, and $B_2$ is fixed at $0.25$ T. (c,d) Simulated H-field amplitudes in the symmetric(a) splitter at $B_1=B_2=0.24$ T and $B_1=B_2=0.22$ T (c), and the asymmetric (b) splitter at $B_1=0.232$ T and $B_1=0.19$ T (d).  The working frequency is fixed at $\omega=1.05\omega_p$. }		
\end{figure}
When two USMP modes are excited simultaneously and interact with each other in the silicon layer, unidirectional SMP multimode interference (USMMI) emerges. To verify this phenomenon, we simulate wave transmission in both symmetric and asymmetric waveguides with the ﬁnite element method (FEM) using COMSOL Multiphysics, as shown in Fig.~\ref{fig:epsart2}(d). In the simulation, we position a magnetic current point source with a frequency of $\omega= 1.05\omega_p$ to excite the SMP and take into account the material loss (i.e., $\nu = 10^{-6} \omega_{p}$). As expected, USMMI is realized at THz frequencies, and it can only propagate forward, not backward, in both symmetric and asymmetric waveguides. The $H$-field distribution in the symmetric waveguide is identical at both InSb-Si interfaces, whereas it differs in the asymmetric waveguide due to structural asymmetry. Moreover, the $H$-fields of USMMI are periodic with a period of 2$L_{\pi}$, where the beat length $L_{\pi}$ is denoted by \cite{SoldanoOpMMI1995,Liu2023LT}
\begin{eqnarray}
	L_{\pi}=\frac{\pi}{k_{even}-k_{odd}}  \label{subeq:9}	 
\end{eqnarray} 
where $k_{even}$ ($k_{odd}$) is the propagation constant of the even (odd) mode, respectively.
\begin{figure*}[t]
	\centering
	\includegraphics[width=0.7\linewidth]{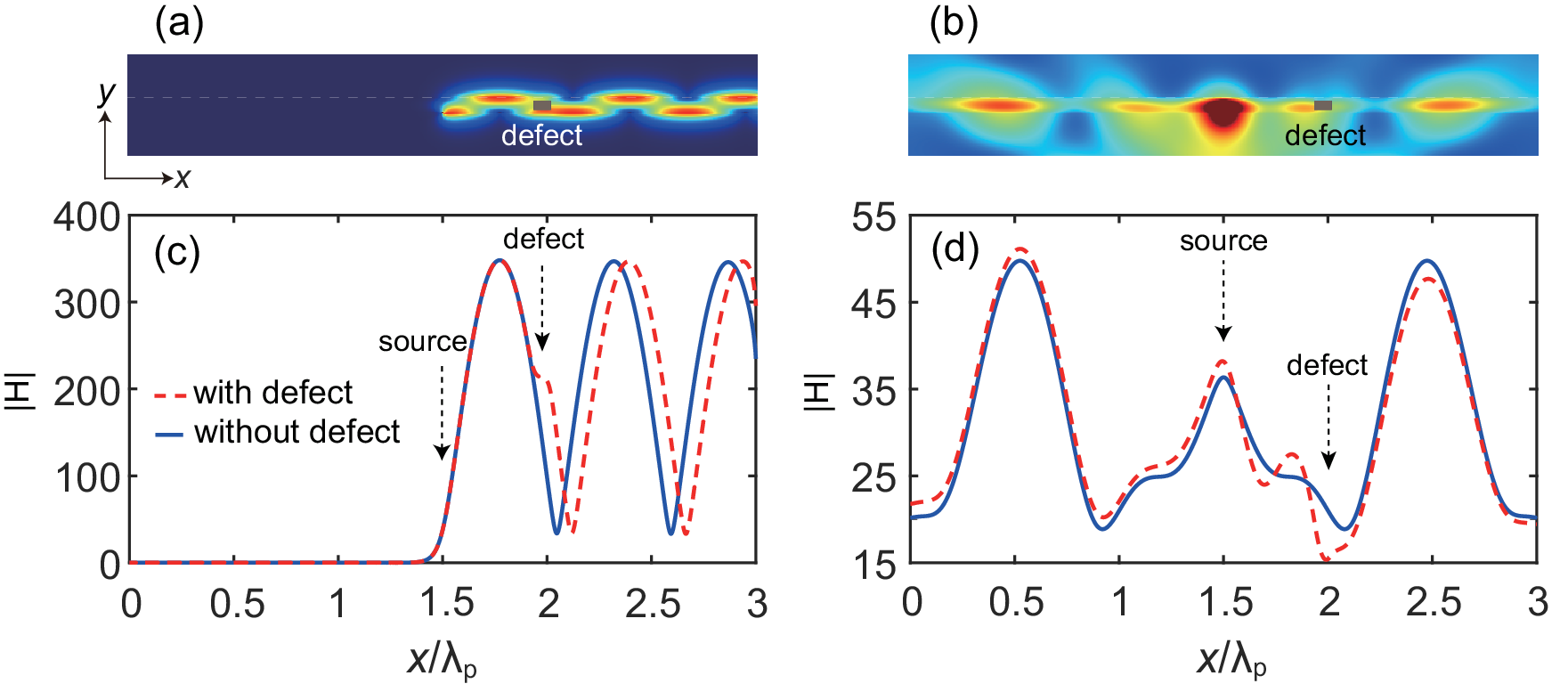}
	\caption{\label{fig:epsart5}Comparison of our unidirectional waveguide with conventional waveguide. (a,b) Simulated H-field amplitudes of the symmetric waveguide. (a) Robust USMP modes without backscattering. (b) A conventional bidirectional mode with backscattering from the metallic defect. (c, d) H-field distributions with (solid lines) and without (dashed lines) the defects along the upper InSb-Si interface. The operating frequency is $\omega = 1.05 \omega_{p}$ and other parameters are the same as in Fig.~\ref{fig:epsart3} }
\end{figure*} 
To further investigate the characteristics of USMMI, we analyze the variation of $L_\pi$ with frequency and the external magnetic field. Figure~\ref{fig:epsart2}(e) shows the analytical value of $L_{\pi}$ as a function of $\omega$ using Eq.~(\ref{subeq:9}) in the whole USMMI range. It is evident that $L_{\pi}$ increases with $\omega$ for both symmetric ($B_1=B_2=0.25$~T) and asymmetric ($B_1=0.15$~T and $B_2=0.25$~T) waveguides, corresponding to the red and blue lines, respectively. Figure~\ref{fig:epsart2}(f) shows $L_{\pi}$ as a function of $B_1$, where $\omega$ is fixed at $1.05\omega_p$ as an example. As seen from the solid red line, the analytical $L_{\pi}$ monotonically decreases with $B_1$ varying from $0.15$~T to $0.25$~T, demonstrating the magnetically-controllable properties of USMMI by tuning two symmetric magnetic fields ($B_1=B_2$). More interestingly, we achieve magnetically-controllable $L_{\pi}$ with a single-sided magnetic field, as indicated by the dashed blue line in Fig.~\ref{fig:epsart2}(f). Here, $B_2$ is fixed at 0.25 T as an example. $L_{\pi}$ monotonically decreases with $B_1$. Furthermore, we select multiple frequencies and magnetic fields to calculate the simulated $L_{\pi}$ with the full-wave FEM, as seen in the squares in Figs.~\ref{fig:epsart2}(e) and~\ref{fig:epsart2}(f), respectively. Obviously, the FEM results are in perfect agreement with the analytical values. Note that USMMI persists in the whole range, resulting from the magnetically controllable USMMI band. This result confirms that USMMI is controllable by both frequency and the external magnetic field.

\subsection{Frequency- and magnetically-tunable arbitrary-ratio power splitter based on USMMI}
 
By controlling the beat length $L_{\pi}$ of MMI with external magnetic fields and frequencies, power manipulation has been achieved in topological PhCs at microwave frequencies \cite{tang2024magnetically}. Here, utilizing USMMI, we achieve arbitrary ratio power manipulation at THz frequencies. To verify this, we propose an H-shaped splitter, as shown in Fig.~\ref{fig:epsart3}(a). The middle part is an SDS waveguide supporting USMMI, while the left and right parts are InSb-Si-Metal waveguides with silicon thickness $d$ \cite{tsakmakidis2017Sci}, which has the same dispersion equation as given in Eq.~(\ref{subeq:4a}). A source, marked by a star, is placed at the input channel S2 to excite USMP and USMMI. Consequently, it is divided into the upper and lower outputs labeled as P1 and P2, respectively. To illustrate this, we perform full-wave simulations in both symmetric ($B_1=B_2=0.25$~T) and asymmetric ($B_1=0.15$~T, $B_2=0.25$~T) waveguides, and define the power ratio of channel P1 (P2) to channel S2 as $\eta_1$($\eta_2$), and the total transmission as $\eta=\eta_1+\eta_2$.

Figure~\ref{fig:epsart3}(b) displays the transmission coefficients $\eta_1$ and $\eta$ versus $\omega$ for both splitters. $\eta_1$ oscillates arbitrarily between 0 and 1 in the unidirectional range from $1.0308\omega_p$ to $1.0778\omega_p$. Owing to the backscattering-immune property of topologically USMP mode and the absence of reflection at the corners, $\eta$ remains close to 1 under low-loss conditions. For the symmetric waveguide, $\eta_1=0$ ($\eta_2=1$) at $\omega=1.041\omega_p$ and $\eta_1=1$ ($\eta_2=0$) at $\omega=1.059\omega_p$ ; whereas for the asymmetric waveguide, $\eta_1=0$ ($\eta_2=1$) at $\omega=1.0335\omega_p$ and $\eta_1=1$ ($\eta_2=0$) at $\omega=1.05\omega_p$. For clarity, the $H$ distributions of both waveguides are presented in Figs.~\ref{fig:epsart3}(c) and~\ref{fig:epsart3}(d), respectively. As expected, the power almost entirely flows into Channels P1 and P2 at $\omega=1.059\omega_p$ and $1.041\omega_p$, and the interference length $L$ satisfies $L\approx 10.5L_{\pi{(\omega=1.059\omega_p)}}\approx 11.5L_{\pi{(\omega=1.041\omega_p)}}$ with an inverted (direct) image at the corner in the symmetric waveguide.  Similarly, the power flow into Channel P1 (P2) at $\omega=1.05\omega_p$ ($1.0335\omega_p$), with $L\approx 8.5L_{\pi{(\omega=1.05\omega_p)}}\approx 9.5L_{\pi{(\omega=1.0335\omega_p)}}$ in the asymmetric waveguide. The results demonstrate that a frequency-tunable arbitrary-ratio power splitter is realized using USMMI.

Furthermore, the magnetically-tunable power splitter based on USMMI is demonstrated. To verify this, the frequency is fixed at $\omega = 1.05 \omega_{p}$. Figure~\ref{fig:epsart4}(a) shows the $\eta_1$, $\eta_2$, and $\eta$ versus $B_1$ by tuning two symmetric magnetic fields ($B_1=B_2$). $\eta_1$ and $\eta_2$ vary continuously from $0$ to $1$ as $B_1$ changes within the range of [0.15 T, 0.25 T]. As a special case, $\eta_1=0$ and $\eta_2=1$ at $B_1=B_2=0.22$ T, and $\eta_1=1$ and $\eta_2=0$ at $B_1=B_2=0.24$ T. The corresponding full-wave simulation results are displayed in Fig.~\ref{fig:epsart4}(c). As expected, the power respectively flows into channel P1 (P2) at $B_1=0.22$ T ($0.24$ T), and the values of $L$ satisfy $L\approx10.5L_{\pi{(B_1=0.24 T)}}\approx9.5L_{\pi{(B_1=0.22 T)}}$ with an inverted (direct) image at the corner. This result confirms the arbitrary ratio power manipulation by the magnetic field. More importantly, we propose a single-sided magnetic control power splitter, which is much easier to operate than the two-sided control in practical applications. To demonstrate this, we set a fixed value of $B_2=0.25$ T as an example and resimulate by only adjusting $B_1$. As seen in Fig.~\ref{fig:epsart4}(b),  the splitter achieves arbitrarily splitting ratios from 0 to 1 as expected. When $B_1=0.232$ T and $B_1=0.19$ T, the power is respectively directed to channels P1 and P2, as shown in Fig.~\ref{fig:epsart4}(d), demonstrating magnetically-controllable power manipulation with only one magnetic field. Therefore, a USMMI-based power splitter has been demonstrated to achieve arbitrary ratios of power manipulation by the external magnetic field intensity and frequency, resulting from the frequency- and magnetically-controllable USMMI.

\subsection{ Robust USMMI at THz frequencies }

To demonstrate the robustness of the USMMI, we introduce a square metallic defect of $0.04\lambda_{p} \times 0.08\lambda_{p}$ into the waveguide, as highlighted by the grey part in Fig.~\ref{fig:epsart5}(a). We take a symmetric waveguide as an example. Figure~\ref{fig:epsart5}(a) shows the simulated $H$-field amplitudes for $\omega = 1.05\omega_{p}$. It can be seen that the excited USMMI mode can smoothly bypass the defect and still travel forward without generating a backward wave from the defect. To clearly illustrate this phenomenon, the $H$-field distributions along the upper InSb-Si interface are displayed for both cases with (dashed line) and without (solid line) defects in Fig.~\ref{fig:epsart5}(c). As expected, the fields of USMMI remain unchanged before the defect and quickly recover the same amplitude after it. These results confirm the strong robustness of USMMI. 

For comparison, we remove the external magnetic field from the waveguide in Fig.~\ref{fig:epsart5}(a), effectively making it a conventional waveguide and resimulating it. Fig.~\ref{fig:epsart5}(b) shows the simulated $H$-field amplitudes with the defect. The excited wave is transmitted bidirectionally, and the H-field radiates everywhere in this waveguide. To clearly show this, the $H$-field distributions for conventional waveguides with and without defects are plotted in Fig.~\ref{fig:epsart5}(d). Without defects, the forward and backward fields are symmetric (solid line). When the defect is introduced, the field amplitudes become asymmetric (dashed line), with an increase before the defect and a decrease after it, demonstrating the backward reflection induced by the defect. These results demonstrate that our waveguide supports a robust USMMI mode without backscattering against defects, in contrast to the conventional bidirectional waveguide with backscattering. 
 
\subsection{ Multiple modes conversion and phase modulation based on USMP}

\begin{figure}[t]
	\centering
	\includegraphics[width=\linewidth]{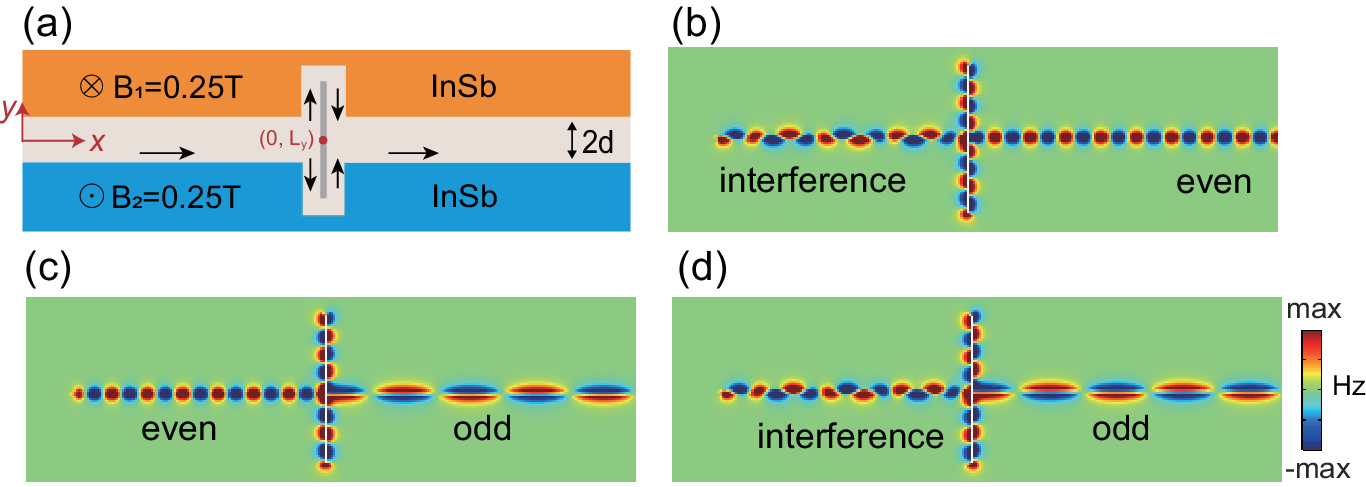}
	\caption{\label{fig:epsart6}(a) Schematic of multiple SMP modes conversion based on the symmetric splitter. Simulated $H_z$ field amplitudes for $\omega = 1.08\omega_{p}$. The interference modes are converted to odd (b) and even (d) modes. (c) The even mode is converted to the odd mode.}
\end{figure}

 Based on the splitter, we further design a structure to achieve the conversion among multiple modes, as shown in Fig.~\ref{fig:epsart6}(a). In this structure, the left and right parts are symmetric waveguides, and the silicon thickness is $d$ in the middle part. The input mode is equally divided and then recombines into a single output mode, with a phase difference $\Delta \varphi=\varphi_0+4kL_y $, where $\varphi_0$ is the initial phase difference, and $L_y$ is the position of the center of the metal bar along the $y$-axis. It is found that the incident mode can be converted to an even mode when $ \Delta \varphi =2 n\pi$ (where n is an integer) and to an odd mode when $ \Delta \varphi =(2n+1)\pi$. The mode conversion satisfies the equations
\begin{align}
	 L_y = \left\{
	\begin{aligned}
		&\frac{2n\pi-\varphi_0}{4k} &  \text{for even mode}, \\
		&\frac{(2n+1)\pi-\varphi_0}{4k}& \text{ for odd mode}.
	\end{aligned}
	\right.\label{subeq:10}
\end{align}
where $n= ...-2, -1, 0, 1, 2... $, this indicates that multimode conversion can be realized by tuning $L_y$. We first consider a special case of mode conversion from an even mode to an odd mode, where $\varphi_0=0$. Here, we take $\omega$=1.08$\omega_p$ as an example, corresponding to $k=2.153k_{p}$. Thus, the theoretical value is found to be $L_y = 0.0580\lambda_{p}$ for $n=0$ using Eq.~(\ref{subeq:10}). To verify this, we perform full-wave simulations and the simulated $H_z$-field for $\omega$=1.08$\omega_p$ is displayed in Fig.~\ref{fig:epsart6}(c). The even mode is converted into odd modes when $L_y = 0.0580\lambda_{p}$ as expected. Furthermore, our FEM simulations demonstrate that the USMMI mode is converted into both even and odd modes when $L_y = -0.0283\lambda_{p}$ and $L_y = 0.0297\lambda_{p}$, as shown in  Figs.~\ref{fig:epsart6}(b) and~\ref{fig:epsart6}(d). It is found that $\varphi_0=0.4882\pi$, and the values of $L_y$ also satisfy Eq. (10) when $n=0$. The results demonstrate that our structure achieves not only the conversion between USMMI and single mode but also the conversion between odd and even modes.
\begin{figure*}[t]
	\centering
	\includegraphics[width=0.70\linewidth]{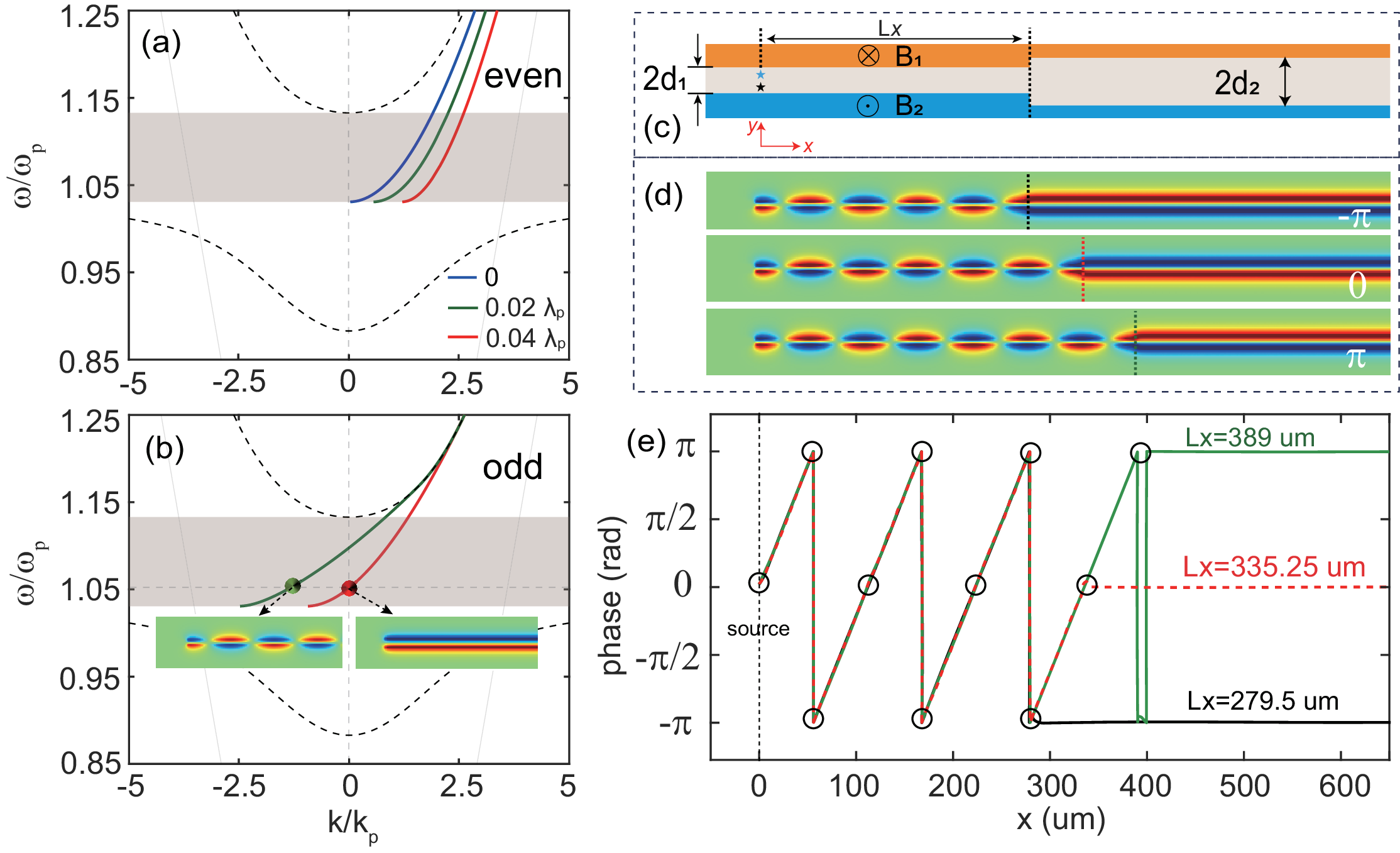}
	\caption{\label{fig:epsart7}(a,b) The dispersion of even and odd modes for different $d$ values in the symmetric SDS waveguide. Insets in (b) show $H_z$ field amplitudes of normal (green dot) and INZ (red dot) mode at $\omega = 1.0517\omega_{p}$. (c)  The schematic of the phase modulator by connecting two SDS waveguides with $d$. (d) Simulated $H_z$ field amplitudes at $\omega = 1.0517\omega_{p}$. (e) The phase of $H_z$ is the output INZ modes along the lower InSb-Si interface for the phase modulation (d). The parameters are $d_1 =0.02 \lambda_{p}$ and $d_2 =0.04 \lambda_{p}$.}
\end{figure*}

Interestingly, we further discover a unique index-near-zero (INZ) odd USMP mode without phase variation ($k=0$) supported by the SDS waveguide at THz frequencies, which is distinct from USMP in the conventional InSb-Si-Metal waveguides \cite{tsakmakidis2017Sci,ZhouINz2022Oe}. These waveguides do not support INZ mode in the upper non-trivial bandgap.  Based on the INZ odd mode in our waveguide, THz phase modulation is achieved. To illustrate this, we first calculate the dispersion curves of odd and even modes for different silicon thicknesses $2d$ in the upper bandgap, as shown in Figs.~\ref{fig:epsart7}(a) and~\ref{fig:epsart7}(b), respectively. Both dispersion curves shift to the left as $d$ decreases. For a given frequency ($\omega$), the propagation constant ($k$) of the even mode gradually approaches zero ($k_{even}\to 0$), but $k_{even}\neq 0$ as $d$ decreases, whose dispersion is the same as that of the conventional InSb-Si-Metal waveguides \cite{tsakmakidis2017Sci}. In contrast to the even mode, the most significant difference from the odd mode is the presence of the INZ mode ($k_{odd}=0$). As an example, the corresponding frequency for the INZ odd mode is $\omega =1.0517\omega_{p}$ when $d=0.04\lambda_{p}$, marked by the red dot in Fig.~\ref{fig:epsart7}(b). Unlike the regular even and odd modes, this INZ odd mode exhibits a stable phase. To clearly verify this, we perform full-wave simulations of the INZ mode at $\omega =1.0517\omega_{p}$, as seen in the right inset of Fig.~\ref{fig:epsart7}(b). For comparison, the result for the regular odd mode when $d=0.02\lambda_{p}$ is displayed in the left inset.  It clearly demonstrates the zero-phase-shift transmission of the INZ odd mode, which is consistent with the theoretical results.

Next, utilizing the INZ odd USMP mode, we design a structure to achieve phase modulation with a phase shift of $2\pi$, as shown in Fig.~\ref{fig:epsart7}(c). This structure consists of two symmetric waveguides with different $d$, and the left part with $2d_1$ supports a non-INZ odd mode, while the right part with $2d_2$ supports an INZ odd mode. The point source (marked by the star) is positioned at a distance of $L_x$ from the interface (dashed line). By adjusting $L_x$,  we can precisely control the output phase ranging from -$\pi$ to $\pi$. To verify this, we perform simulations for different $L_x $ values at $\omega =1.0517\omega_{p}$. As an example, $d_1=0.02\lambda_{p}$ and $d_2=0.04\lambda_{p}$, corresponding to the green and red dots. Figure~\ref{fig:epsart7}(d) shows the simulated $H_z$ field amplitudes for $L_x = 279.5$ um, $325.25$ um, and $389$ um, with the corresponding output phases being -$\pi$, 0, and $\pi$.  To more clearly demonstrate this, we plot the phase of the $H_z$ field along the lower InSb-Si interface for these three $L_x$ values, as shown in Fig.~\ref{fig:epsart7}(e). As expected, the output phases are -$\pi$, 0, and $\pi$ due to the ’super-coupling’ effect \cite{EngNearZero2013,liu2025OL}. These results show that our waveguide can realize all-optical phase modulation with a phase shift ranging from -$\pi$ to $\pi$.

\section{Discussion}
\begin{figure}[t]
	\centering
	\includegraphics[width=\linewidth]{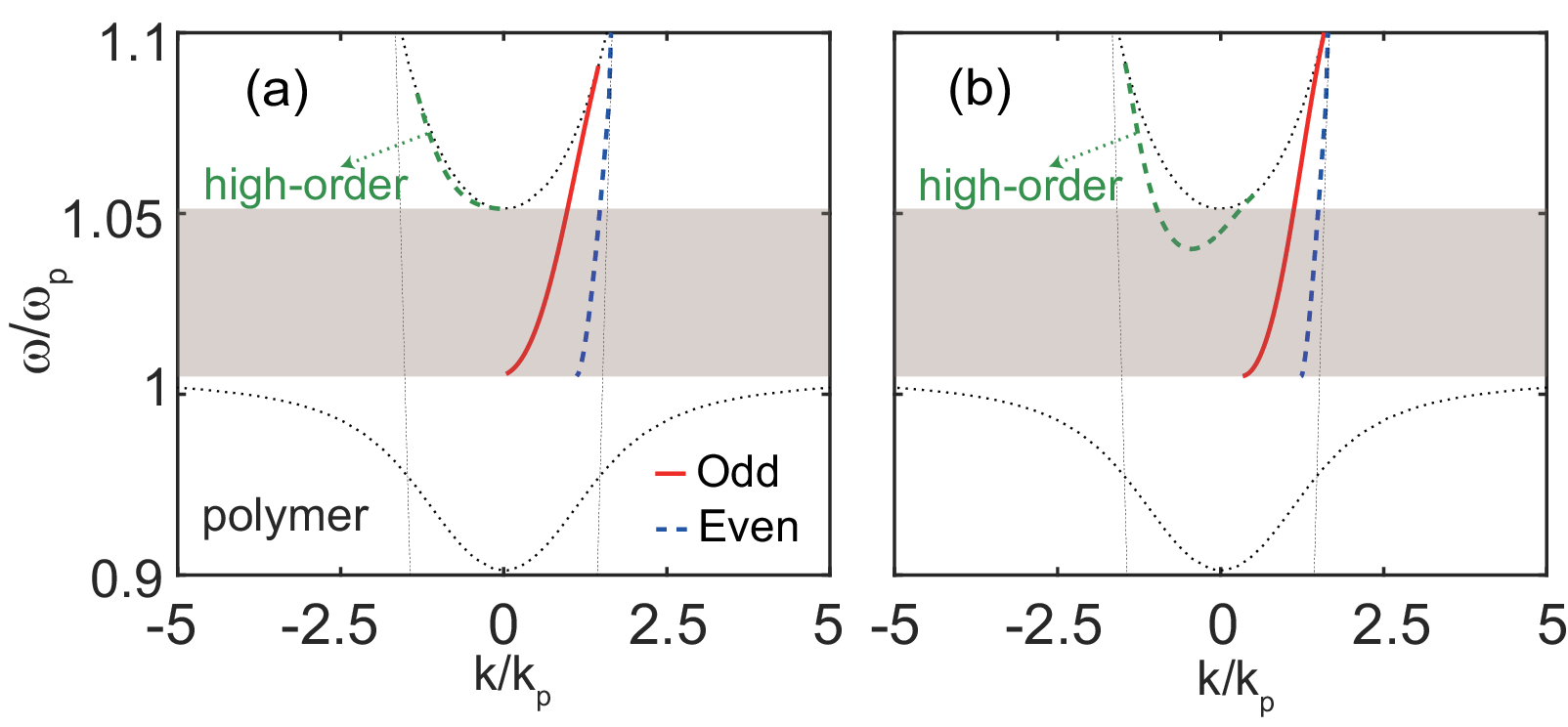}
	\caption{\label{fig:epsart8}The dielectric layer is replaced by polymer ($\varepsilon_{r}=2.28$) instead of Si ($\varepsilon_{r}=11.68$). (a, b) Dispersion relations of SMPs for the symmetric ($\omega_{c_1}=\omega_{c_2}=0.1\omega_p$) waveguide. (a) $d=0.154\lambda_{p}$ and (b) $d=0.2\lambda_{p}$. The blue, red, and green lines show the even, odd, and high-order SMP modes, respectively.}
\end{figure}

Note that the performance of the SMP mode is strongly dependent on the waveguide parameters, such as thickness, dielectric constant of the dielectric layer, and the applied magnetic field. The effects of dielectric thickness and magnetic field strength are previously illustrated in Fig. \ref{fig:epsart2}, using silicon (Si) as a representative material. Here, we further analyze the effect of the dielectric constant on the waveguide characteristics by replacing Si ($\varepsilon_{r}=11.68$) with a polymer ($\varepsilon_{r}=2.28$). Numerical calculations of the dispersion relations are performed for a symmetric waveguide with $\omega_{c_1}=\omega_{c_2}=0.1\omega_{p}$. Figure \ref{fig:epsart8}(a) shows the dispersion curves for $d=d_c=0.154\lambda_{p}$, where $d_{c} = \displaystyle\pi c/(2 \sqrt{\varepsilon_{r}} \omega_{b_2})$ is the critical value $d$ of the polymer. Evidently, two USMP modes (red and blue lines) are supported in the upper bulk mode bandgap [1.005$\omega_{p}$,1.0513$\omega_{p}$]. When $d= 0.2\lambda_{p}>d_c$, the waveguide supports three SMP modes in the upper bandgap: unidirectional even and odd SMP modes, and a bidirectional high-order SMP mode, as shown in Fig. \ref{fig:epsart8}(b). These results are consistent with those presented in Fig. 2. Furthermore, replacing Si with a polymer, the critical thickness $2d_c$ of the dielectric layer is increased from $0.128\lambda_{p}$ to $0.308\lambda_{p}$. It is noteworthy that the thickness can be further extended to 4$\lambda_{p}$ by employing a uniaxial $\varepsilon$-near-zero (UENZ) material, as described in ref. \cite{shen2023SR}. The results indicate that our waveguide supports two USMP modes for different dielectric materials.
\begin{figure}[t]
	\centering
	\includegraphics[width=\linewidth]{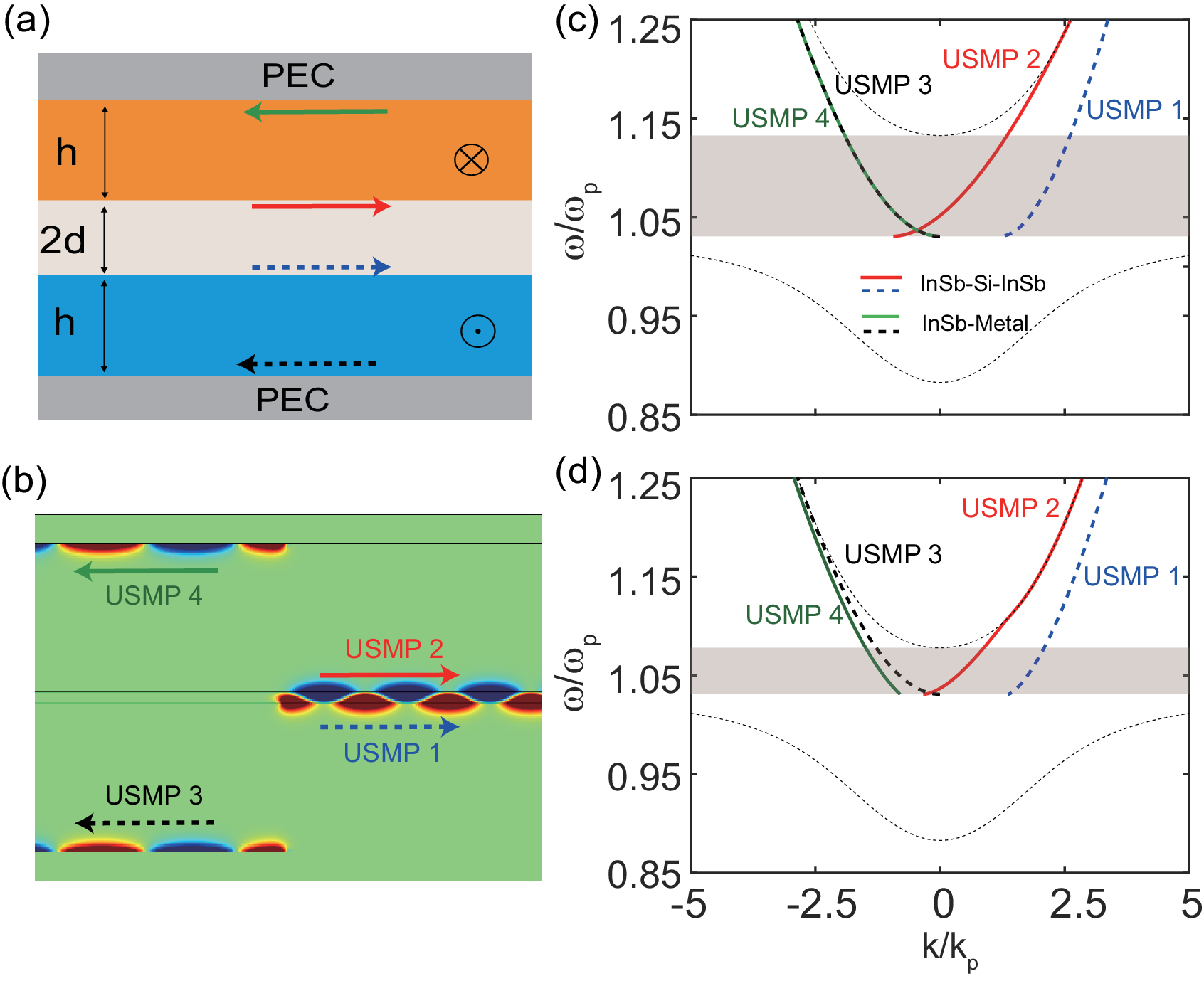}
	\caption{\label{fig:epsart9}Multiple USMP modes. (a) Schematic of the Metal-InSb-Si-InSb-Metal waveguide. (b) Simulated $H_z$ field amplitudes for $\omega = 1.05\omega_{p}$. (c, d) Dispersion relations of SMPs for the symmetric ($\omega_{c_1}=\omega_{c_2}=0.25\omega_p$) and the asymmetric waveguide ($\omega_{c_1} = 0.15\omega_p$ and $\omega_{c_2} = 0.25\omega_p$) in the upper bulk mode bandgap. The other parameters are $h=\lambda_{p}$ and $d=0.04\lambda_{p}$}.
\end{figure} 

Furthermore, more topological USMP modes can be supported by introducing metallic truncations to our InSb-Si-InSb waveguide, as the InSb-Metal interface supports USMP mode \cite{hassani2019truly}. To verify this, we calculated the dispersion relations of the waveguide shown in Fig. \ref{fig:epsart9}(a). Figures \ref{fig:epsart9}(c) and \ref{fig:epsart9}(d) illustrate the dispersion curves for symmetric and asymmetric waveguides with $d=0.04\lambda_{p}$ and $h=\lambda_{p}$. Clearly, both waveguides support four USMP modes:  the InSb-Si interfaces support USMP1 and USMP2, while the upper and lower InSb-Metal interfaces support USMP4 and USMP3, respectively. For clarity, we simulated the wave propagation at $\omega = 1.05\omega_{p}$ in the symmetric waveguide, as shown in Fig. \ref{fig:epsart9}(b). As expected, the middle InSb-Si-InSb structure supports forward USMMI based on USMP1 and USMP2, while the upper and lower InSb-Metal interfaces support backward USMP propagation. These results demonstrate that multiple topological modes can be realized by introducing metallic truncation to our system.

\section{Conclusion}  
In this work, we have proposed and investigated a slab waveguide consisting of a non-gyroelectric dielectric layer sandwiched between two gyroelectric semiconductors subjected to opposite external magnetic fields. The dispersion characteristics of SMPs are analytically derived in both local and non-local models. Our numerical results show that the waveguide supports two truly USMP modes in the upper topological bulk mode bandgap. However, in the lower bulk mode bandgap, SMPs lose their strict unidirectionality due to non-local effects. 
By leveraging the two USMP modes, we demonstrate a magnetically-controllable USMMI at THz frequencies. Notably, USMMI exhibits strong robustness against defects, with no backscattering, in contrast to conventional bidirectional waveguides, which are susceptible to backscattering. Furthermore, we realized a frequency- and magnetically-tunable arbitrary-ratio splitter based on robust USMMI, enabling precise control of the splitter's properties.
Additionally, multimode conversion is achieved based on the splitter. We also identify a unique index-near-zero (INZ) odd USMP mode supported by our waveguide, distinct from conventional InSb-Si-Metal waveguides. Utilizing the INZ mode with zero phase shift transmission, we have designed a phase modulator that precisely controls the phase from -$\pi$ to $\pi$. These results in mode manipulation utilizing USMP - encompassing interference, power, and phase control - offer a novel approach for the flexible manipulation of THz topological waves.

\begin{acknowledgments}
This work was supported by National Natural Science Foundation of China (12464057, 12104203, 12264027, 61927813); Natural Science Foundation of Jiangxi Province (20242BAB25039, 20224BAB211015, 20242BAB20030, 20242BAB20024); Jiangxi Double-Thousand Plan (jxsq2023101069); and the Research Program of NUDT (ZK22-17). K. L. Tsakmakidis acknowledges funding within the
framework of the National Recovery and Resilience Plan Greece 2.0, funded by the European Union NextGenerationEU (Implementation body: HFRI), under Grant No. 16909. K. L. Tsakmakidis was also supported by the General Secretariat for Research and Technology and the Hellenic Foundation for Research and Innovation under Grant No. 4509.
\end{acknowledgments}

\appendix
\nolinenumbers\section{Derivation of dispersion formula in local model}
In this appendix, we demonstrate Eq.~(\ref{subeq:3}) in the main text. In the local model,  the magnetic field
of SMP has non-zero components 
\begin{equation}
	\begin{aligned}
		& H_{z_1}=Ae^{-\alpha_1 y}e^{i(kx-\omega t)},                           & y\geq d\\   
		&{H_z}=\!\!(G_1e^{\alpha_d y}\!\!+\!G_2e^{-\alpha_d y})e^{i(kx-\omega t)},  &  -d < y < d\\    
		& H_{z_2}=Ce^{\alpha_2 y}e^{i(kx-\omega t)},                            & y\leq -d    
	\end{aligned}\label{subeq:A1}   
\end{equation}
for the upper semiconductor layer, middle dielectric layer, and lower semiconductor layer, respectively. Using Maxwell’s equations $\nabla\times \textbf{H}=-i\omega\varepsilon_0 \varepsilon_\infty\textbf{E}$ and $\nabla\times \textbf{E}=i\omega\mu_0 \textbf{H}$, the non-zero components ($ E_x$ and $E_y$) of the electric field in each layer can be directly derived from $H_z$ in Eq.~(\ref{subeq:A1}), thus $E_x$ can be written as
\begin{equation}
	\begin{aligned}
		& E_{x_1}= -\frac{(i\alpha_1\varepsilon_{1_1}-ik\varepsilon_{2_1}) }{\omega \varepsilon_0 \varepsilon_{v_1}\varepsilon_{1_1} }   Ae^{-\alpha_1y}e^{i(kx-\omega t)}, &y\geq d\\	
		& {E_x}= -\frac{\alpha_d}{i\omega\varepsilon_0\varepsilon_r}(G_1e^{\alpha_d y}-G_2e^{-\alpha_d y})e^{i(kx-\omega t)},&-d<y<d\\
		& E_{x_2}= \frac{(i\alpha_2\varepsilon_{1_2}-ik\varepsilon_{2_2}) }{\omega \varepsilon_0 \varepsilon_{v_2}\varepsilon_{1_2} } Ce^{-\alpha_2 y}e^{i(kx-\omega t)}, &y\leq -d   
	\end{aligned}\label{subeq:A2}
\end{equation}
According to the boundary conditions of fields, $H_z$ and $E_x$ are continuous at boundaries $y = -d$ and $y =d$. Considering the continuity of $ H_z$, which requires 
$H_{z_1}\mid_{y=d}={H}_z\mid_{y=d}$ and $H_{z_2}\mid_{y=-d}={H}_z\mid_{y=-d}$, we obtain from Eq.~(\ref{subeq:A1})
\begin{equation}
	\begin{aligned}
		Ae^{-\alpha_1d}&=\!\!G_1e^{\alpha_d d}\!\!+\!G_2e^{-\alpha_d d}\\
		Ce^{-\alpha_2 d}&=\!\!G_1e^{-\alpha_d d}\!\!+\!G_2e^{\alpha_d d}  
	\end{aligned}\label{subeq:A3}
\end{equation}
Considering the continuity of $E_x$, which satisfies $E_{x_1}\mid_{y=d}={E}_x\mid_{y=d}$ and $E_{x_2}\mid_{y=-d}={E}_x\mid_{y=-d}$, we obtain from Eq.~(\ref{subeq:A2})
\begin{equation}
	\begin{aligned}
		(k\frac{\varepsilon_{2_1} }{ \varepsilon_{1_1} }-\alpha_1)   Ae^{-\alpha_1 d}&=\varepsilon_{v_1}\frac{\alpha_d}{\varepsilon_r}(G_1e^{\alpha_d d}-G_2e^{-\alpha_d d})\\
		(k\frac{\varepsilon_{2_2} }{ \varepsilon_{1_2} }-\alpha_2) Ce^{\alpha_2 d}&=\varepsilon_{v_2}\frac{\alpha_d}{\varepsilon_r}(G_2e^{\alpha_d d}-G_1e^{-\alpha_d d})\\ 
	\end{aligned}\label{subeq:A4}
\end{equation}
By eliminating the four coefficients $A$, $C$, $G_1$ and $G_2$ in Eqs.~(\ref{subeq:A3}) and~(\ref{subeq:A4}), the dispersion relation of SMPs can be derived as
\begin{align}
	e^{4\alpha_d d}=\frac{[1-\displaystyle\frac{ \varepsilon_{r}(\varepsilon_{1_1}\alpha_1-k\varepsilon_{2_1})}{\varepsilon_{1_1}\alpha_d\varepsilon_{v_1}}][1-\frac{\varepsilon_{r}(\varepsilon_{1_2}\alpha_2-k\varepsilon_{2_2})}{\varepsilon_{1_2}\alpha_d\varepsilon_{v_2}}]}{[1+\displaystyle\frac{ \varepsilon_{r}(\varepsilon_{1_1}\alpha_1-k\varepsilon_{2_1})}{\varepsilon_{1_1}\alpha_d\varepsilon_{v_1}}][1+\frac{\varepsilon_{r}(\varepsilon_{1_2}\alpha_2-k\varepsilon_{2_2})}{\varepsilon_{1_2}\alpha_d\varepsilon_{v_2}}]}
\end{align}
which corresponds to Eq.~(\ref{subeq:3}) of the main text.

\section{Derivation of formula in non-local model}
In this appendix, we demonstrate Eq.~(\ref{subeq:6}) in the main text. When non-local effects are taken into account, the dispersion relation of SMP can be derived by solving the hydrodynamic and Maxwell’s equations as follows \cite{hassani2019truly}
\setlength{\belowdisplayskip}{10pt}
\begin{equation}
	\begin{aligned}	
		&\nabla\times \textbf{H}=-i\omega\varepsilon_0 \varepsilon_\infty\textbf{E} +\textbf{J}\\ &\nabla\times \textbf{E}=i\omega\mu_0 \textbf{H}\\ 
		&\beta^2\nabla(\nabla\cdot \textbf{J})+\omega(\omega+i\nu)\textbf{J}+i\omega\textbf{J} \times\omega_{c}\hat z=i\omega\omega_p^2\varepsilon_{0}\varepsilon_\infty \textbf{E} \\
	\end{aligned}
\end{equation}
Since the SDS waveguide only supports the TM mode ($H_x = H_y = E_z = 0$), the above equations for the lossless case ($\nu=0$) can be expressed as
\setlength{\belowdisplayskip}{10pt} 
\begin{equation}
	\begin{aligned}			
		&	\frac{\partial \widetilde{E}_{y}}{\partial x}-\frac{\partial \widetilde{E}_{x}}{\partial y}=i \omega \mu_0 \widetilde{H}_{z} \\
		&	\frac{\partial \widetilde{H}_{z}}{\partial y}=-i\omega\varepsilon_0 \varepsilon_\infty \widetilde{E}_{x}+\widetilde{J}_{x}\\
		&	-\frac{\partial \widetilde{H}_{z}}{\partial y}=-i\omega\varepsilon_0 \varepsilon_\infty \widetilde{E}_{y}+\widetilde{J}_{y}\\
		&	\beta^2[\frac{\partial^2 \widetilde{J}_{x}}{\partial x^2}+\frac{\partial^2 \widetilde{J}_{y}}{\partial x\partial y}] +\omega^2 \widetilde{J}_{x} +i\omega  \omega_c \widetilde{J}_{y}=i\omega {\omega_p}^2\varepsilon_0 \varepsilon_\infty \widetilde{E}_{x}\\
		&	\beta^2[\frac{\partial^2 \widetilde{J}_{x}}{\partial x\partial y}+\frac{\partial^2 \widetilde{J}_{y}}{\partial y^2}] +\omega^2 \widetilde{J}_{y} -i\omega  \omega_c \widetilde{J}_{x}=i\omega {\omega_p}^2\varepsilon_0 \varepsilon_\infty \widetilde{E}_y\\	
	\end{aligned}
\end{equation}
In the non-local model, the non-zero field components ($\widetilde{E}_x, \widetilde{E}_y, \widetilde{H}_z$) and the normal component ($\widetilde{J}$) are found to have the form \cite{yan2023broadband}
\begin{equation}\left.
	\begin {aligned}
	&	\widetilde{E}_{x_1}=(A_1\exp^{ip_{1}y}+\!A_1' e^{-\gamma_{1}y})e^{i(kx\!-\!\omega t)}\\
	&   \widetilde{E}_{y_1}=-i(s_1 A_1e^{ip_{1}y}+ s_1' A_1'e^{-\gamma_{1}y})e^{i(kx-\omega t)}\\
	&	\widetilde{H}_{z_1}=\frac{i}{\omega\mu_0}(L_1 A_1e^{ip_{1}y}+L_1' A_1'e^{-\gamma_{1}y})e^{i(kx-\omega t)}\\
	&	\widetilde{J}_{y_1}=-ik\widetilde{H}_{z_1}+i\omega\varepsilon_0 \varepsilon_\infty \widetilde{E}_{y_1}\\
	\end {aligned}
	\right.\label{subeq:B3}
	\quad y\geq d
\end{equation} 
for the upper semiconductor, and they can be written as
\begin{equation}\left.
	\begin{aligned}
		&	\widetilde{E}_{x_2}=(A_2e^{-ip_{2}y}+A_2'e^{\gamma _{2}y})e^{i(kx-\omega t)}  \\		
		&	\widetilde{E}_{y_2}=i(s_2 A_2e^{-ip_{2}y}+ s_2' A_2'e^{\gamma_{2} y})e^{i(kx-\omega t)}\\
		&	\widetilde{H}_{z_2}=-\frac{i}{\omega\mu_0}(L_2 A_2e^{-ip_{2}y}\!\!+\!L_2' A_2' e^{\gamma_{2} y})e^{i(kx\!-\!\omega t)}\\
		&	\widetilde{J}_{y_2}=-ik\widetilde{H}_{z2}+i\omega\varepsilon_0 \varepsilon_\infty \widetilde{E}_{y2}\\
	\end{aligned}\label{subeq:B4}
	\right.
	\quad y\leq -d
\end{equation}
for the lower semiconductor, the parameters are given in the main text. According to the continuous boundary conditions of electric fields, we have $\widetilde	{E}_{x_1}\mid_{y=d}={E}_x\mid_{y=d}$ and $\widetilde	{E}_{x_2}\mid_{y=-d}={E}_x\mid_{y=-d}$. Combining this with the first equations in Eq.~(\ref{subeq:B3}) and Eq.~(\ref{subeq:B4}) and the second equations in Eq.~(\ref{subeq:A2}), it is found that
\begin{equation}
	\begin{aligned}
		A_1e^{ip_{1}d}+\!A_1'e^{-\gamma_{1}d}&=-\frac{\alpha_d}{i\omega\varepsilon_0\varepsilon_r}(G_1e^{\alpha_d d}-G_2e^{-\alpha_d d})\\
		A_2e^{ip_{2}d}+A_2'e^{-\gamma _{2}d}&=-\frac{\alpha_d}{i\omega\varepsilon_0\varepsilon_r}(G_1e^{-\alpha_d d}-G_2e^{\alpha_d d})  
	\end{aligned}\label{subeq:B5}
\end{equation}
According to the continuous boundary conditions of magnetic fields, $\widetilde	{H}_{z_1}\mid_{y=d}={H}_z\mid_{y=d}$ and $\widetilde	{H}_{z_2}\mid_{y=-d}={H}_z\mid_{y=-d}$. By combining the third equation in Eqs.~(\ref{subeq:B3}) and~(\ref{subeq:B4}) with the second equation in Eq.~(\ref{subeq:A1}), we obtain
\begin{equation}
	\begin{aligned}
		\frac{i}{\omega\mu_0}(L_1 A_1e^{ip_{1}d}+L_1' A_1'e^{-\gamma_{1}d})&=\!\!G_1e^{\alpha_d d}\!\!+\!G_2e^{-\alpha_d d}\\
		-\frac{i}{\omega\mu_0}(L_2 A_2e^{ip_{2}d}\!\!+\!L_2' A_2' e^{-\gamma_{2} d})&=\!\!G_1e^{-\alpha_d d}\!\!+\!G_2e^{\alpha_d d}  
	\end{aligned}\label{subeq:B6}
\end{equation}
Unlike the local model, an additional boundary condition for the non-local hydrodynamic model is required at the magnetized InSb-Si interface ($x$-$z$ plane): $\textbf{J} \cdot \hat y=0 $ \cite{buddhiraju2020absence}. This is because the theoretical basis of this model is based on the assumption that the electron density sharply vanishes perpendicular to the interface. Thus, the additional boundary condition can be written as $\widetilde{J}_{y_1}\mid_{y=d}=0$ and $\widetilde{J}_{y_2}\mid_{y=-d}=0$.  Using the last equations from Eq.~(\ref{subeq:B3}) and Eq.~(\ref{subeq:B4}), we have
\begin{equation}
	\begin{aligned}
		R_1A_1e^{ip_{1}d} +R_1'A_1'e^{-\gamma_1 d}&=0\\
		R_2A_2e^{ip_{2}d} +R_2'A_2'e^{-\gamma_2 d}&=0
	\end{aligned} \label{subeq:B7}
\end{equation}
where $R_j=\displaystyle\frac{k}{\omega\mu_0}L_j +\omega\varepsilon_0 \varepsilon_\infty s_j$ and $R_j'=\displaystyle\frac{k}{\omega\mu_0}L_j' +\omega\varepsilon_0 \varepsilon_\infty s_j'$, with $L_j=-\!k s_j\!\!+\!ip_{j}$ and $L_j'=-\!k s_j'\!-\!\!\gamma_{j}$, $j=1,2$. By eliminating the six coefficients $A_1$, $A_2$, $A_1'$, $A_2'$, $G_1$ and $G_2$ in Eqs.~(\ref{subeq:B5})- Eq.~(\ref{subeq:B7}), the dispersion relation of SMPs can be derived as
\begin{eqnarray}
	e^{4\alpha_d d}=\prod_{j=1}^{2} N_j
\end{eqnarray}
with
\begin{eqnarray*}
	\!\!N_j=\frac{\varepsilon_\infty\alpha_d/\varepsilon_r(\gamma_{j}s_j\!\!+\!\! ip_{j}s_j')\!\!+\!\!k(\gamma_{j}\!\!+\!\!ip_{j})\!\!+\!\!(k^2\!\!-\!\!k_{0}^2\varepsilon_\infty)(s_j'\!\!-\!\! s_j)}{\varepsilon_\infty\alpha_d/\varepsilon_r(\gamma_{j}s_j\!\!+\!\!ip_{j}s_j')\!\!-\!\!k(\gamma_{j}\!\!+\!\!ip_{j})\!\!-\!\!(k^2\!\!-\!\!k_0^2\varepsilon_\infty)(s_j'\!\!-\!\! s_j)}\!\!
\end{eqnarray*}
which corresponds to Eq.~(\ref{subeq:6}) of the main text.

\bibliography{apssamp_Prb}% Produces the bibliography via BibTeX.

\end{document}